\documentclass[acmtog,authorversion]{acmart}
\acmSubmissionID{405}

\usepackage{booktabs} 
\usepackage{subcaption}

\usepackage{verbatim}

\usepackage{flafter}
\usepackage{placeins}
\usepackage{footmisc}

\newcommand{\andy}[1]{\footnote{\textcolor{blue}{#1}}}
\usepackage{dsfont}
\usepackage[ruled]{algorithm2e} 

\SetAlFnt{\small}
\SetAlCapFnt{\small}
\SetAlCapNameFnt{\small}
\SetAlCapHSkip{0pt}

\setcopyright{rightsretained}

\begin{document}

\acmJournal{TOG}
\acmYear{2019}\acmVolume{38}\acmNumber{4}\acmArticle{65}\acmMonth{7} \acmDOI{10.1145/3306346.3323020}

\sloppy

\title{Neural Volumes: Learning Dynamic Renderable Volumes from Images}



\author{Stephen Lombardi}
\affiliation{%
 \institution{Facebook Reality Labs}
 }
\email{stephen.lombardi@fb.com}

\author{Tomas Simon}
\affiliation{%
 \institution{Facebook Reality Labs}
 }
\email{tomas.simon@fb.com}

\author{Jason Saragih}
\affiliation{%
 \institution{Facebook Reality Labs}
 }
\email{jason.saragih@fb.com}

\author{Gabriel Schwartz}
\affiliation{%
 \institution{Facebook Reality Labs}
 }
\email{gabe.schwartz@oculus.com}

\author{Andreas Lehrmann}
\affiliation{%
 \institution{Facebook Reality Labs}
 }
\email{asml@fb.com}

\author{Yaser Sheikh}
\affiliation{%
 \institution{Facebook Reality Labs}
 }
\email{yasers@fb.com}


\renewcommand{\shortauthors}{Lombardi et al.}

\begin{teaserfigure}
    \centering
    \includegraphics[width=1.0\textwidth]{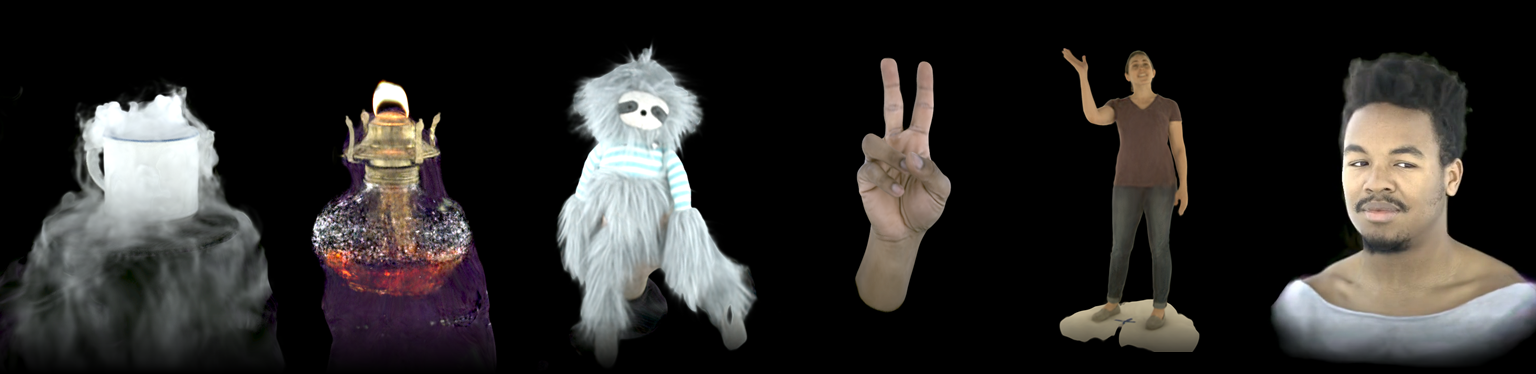}
    \caption{Renderings of objects captured and modeled by our system. The input to our method consists of synchronized and calibrated multi-view video. We build a dynamic, volumetric representation of the scene by training an encoder-decoder network end-to-end using a differentiable ray marching algorithm.}
    \label{fig:teaser}
\end{teaserfigure}

\begin{abstract}
Modeling and rendering of dynamic scenes is challenging, as natural scenes often contain complex phenomena such as thin structures, evolving topology, translucency, scattering, occlusion, and biological motion. Mesh-based reconstruction and tracking often fail in these cases, and other approaches (e.g., light field video) typically rely on constrained viewing conditions, which limit interactivity. We circumvent these difficulties by presenting a learning-based approach to representing dynamic objects inspired by the integral projection model used in tomographic imaging. The approach is supervised directly from 2D images in a multi-view capture setting and does not require explicit reconstruction or tracking of the object. Our method has two primary components: an encoder-decoder network that transforms input images into a 3D volume representation, and a differentiable ray-marching operation that enables end-to-end training. By virtue of its 3D representation, our construction extrapolates better to novel viewpoints compared to screen-space rendering techniques. The encoder-decoder architecture learns a latent representation of a dynamic scene that enables us to produce novel content sequences not seen during training. To overcome memory limitations of voxel-based representations, we learn a dynamic irregular grid structure implemented with a warp field during ray-marching. This structure greatly improves the apparent resolution and reduces grid-like artifacts and jagged motion. Finally, we demonstrate how to incorporate surface-based representations into our volumetric-learning framework for applications where the highest resolution is required, using facial performance capture as a case in point.
\end{abstract}

%
%


\maketitle

\section{Introduction}
Polygon meshes are an extremely popular representation for 3D geometry in photo-realistic scenes.
Mesh-based representations efficiently model solid surfaces and can be paired with sophisticated reflectance functions to generate compelling renderings of natural scenes.
In addition, there has been significant progress recently in optimization techniques to support real-time ray-tracing, allowing for interactivity and immersion in demanding applications such as Virtual Reality (VR).
However, little of the interactive photo-real content available today is data-driven because many real-world phenomena are challenging to reconstruct and track with high fidelity.
State-of-the-art motion capture systems struggle to handle complex occlusions (e.g., running hands through one's hair), to account for reflectance variability (e.g., specularities in the sheen of a moving object), or to track topological evolution in dynamic participating media (e.g., smoke as it billows upward). Where solutions exist, they are typically specialized to an individual phenomenon~\cite{Xu:ToG:2014,Atcheson:ToG:2008,Roth:CVPR:2006,Hawkins:2005}, and are often aimed at either image generation \cite{Buehler:2001,Kalantari:2016} or 3D reconstruction \cite{Goesele:2007,Niessner:2013}, but not both.
Since mesh-based representations rely heavily on the quality of reconstruction to produce compelling renderings, they are ill-suited to handle such cases.
Nonetheless, these kinds of phenomena are necessary to create compelling renderings of much of our natural world.


To address limitations posed by inaccurate geometric reconstructions, great progress has been made in recent years by relaxing the physics-inspired representation of light transport, and instead leveraging machine learning to bridge the gap between the representation and observed images of the scene. In~\citet{Lombardi:2018}, this technique was used to great effect in modeling the human face, where it was demonstrated that a neural network can be trained to compensate for geometric reconstruction and tracking inaccuracies through a view-dependent texture. Similar approaches have also been shown to be effective for modeling 
general far-field scenes~\cite{Overbeck:2018}. An extreme variant of this technique is screen-space rendering, where no geometry of the scene is used at all~\cite{Karras:2018,Wang:2018}. Although these approaches have been shown to produce high quality renderings of complex scenes, they are limited to viewpoints available to the system at training time. Since their neural architectures are not \emph{3D aware}, the methods do not extrapolate to novel viewpoints in a way that is consistent with the real world. The problem is exacerbated when modeling near-field scenes, where variation in viewpoint is more common as a user interacts with objects in the scene, compared with far-field captures where there is less interactivity and the viewer is mainly stationary. 

An important insight in this work is that if \emph{both} geometry and appearance variations can be learned simultaneously, phenomena explainable by geometric variations may be modeled as such, leading to better generalization across viewpoints.
The challenge, then, is to formulate this joint learning such that good solutions can be found. Directly optimizing over a mesh-representation using gradient-based optimization is prone to terminating in poor local minima. This is the case even when a model of both appearance and geometry are known \emph{a priori}, and is exacerbated when these models are also unknown.
One of the main reasons for this difficulty is the local support of the gradients of mesh-based representations. To address this, we propose using a volumetric representation consisting of opacity and color at each position in 3D space, where rendering is realized through integral projection. During optimization, this semi-transparent representation of geometry disperses gradient information along the ray of integration, effectively widening the basin of convergence, enabling the discovery of good solutions. 

Although the volumetric representation has the ability to represent 3D geometric phenomena in a geometrically faithful way, it can easily over-fit from image-supervision for a typical density of viewpoints. As such, additional regularization is necessary to achieve good results. In this work, we show that a neural-network decoder is sufficient to encourage discovery of solutions that generalize across viewpoints. At first glance, this may appear surprising given the decoder network typically has enough capacity to reproduce solutions found through a direct solve of the volume's entries. We conjecture that the decoder network introduces spatial regularity into the gradients of the volume's entries (i.e., opacity and color), leading to more generalizable solutions without diminishing the volumetric representation's capacity.
Additionally, the decoder network is paired with an encoder network that produces a low-dimensional latent space that encodes the state of the scene at each frame, enabling joint reconstruction of sequences rather than just individual frames.
Analogous to non-rigid structure from motion~\cite{Torresani:2008}, this architecture can leverage a scene's regularity across time to improve viewpoint generalization further.
This latent code can be used to generate novel renderings of the scene's content by traversing the latent space, enabling realistic modifications of the recording, or even completely new sequence animation, without requiring object/scene/content specific solutions.

Despite these advantages of the volumetric representation, its main drawback is limited resolution. Using voxel-based data structures to represent a scene typically requires an order of magnitude more memory than its mesh-based counterpart to achieve similar levels of resolution. Furthermore, much of this memory is dedicated to modeling empty space or the inside of objects; neither of which have an impact on the rendered result. This limitation stems from the regular grid structure this representation exhibits. To overcome this limitation, we employ a warping technique that indirectly escapes the restrictions imposed by a regular grid structure, allowing the learning algorithm to make the best use of available memory. Using this technique, we demonstrate significantly higher fidelity than using only a conventional voxel data structure. Furthermore, as our representation is 3D based, we can naturally combine it with surface-based  reconstruction and tracking methods when appropriate. This allows us to reach the highest levels of fidelity on objects in the scene for which state-of-the-art reconstruction and tracking work well, while also maintaining a complete model of the scene. 

In summary, we propose a novel volumetric representation that is object/scene agnostic, can generalize well to novel viewpoints, reconstructs dynamic scenes jointly, facilitates novel content generation, requires only image-level supervision and is end-to-end trainable. The resulting models afford real-time rendering and support on-the-fly adjustments, suitable for interactive applications in VR. In \S\ref{sec:related_work} we cover related work, followed by an overview of our approach in \S\ref{sec:overview}. Details of the encoder and decoder architectures are covered in sections \S\ref{sec:encoder} and \S\ref{sect:volume_decoders} respectively. Rendering through integral projection is discussed in \S\ref{sect:rendering} and details of the learning problem are covered in \S\ref{sec:training}. We evaluate this architecture on a number of challenging scenes and present ablation study on various design choices in our construction in \S\ref{sec:Experiments}. We conclude in \S\ref{sec:discussion} with a discussion and directions of future work.

\begin{figure*}[t]
    \begin{center}
    \includegraphics[width=0.99\textwidth]{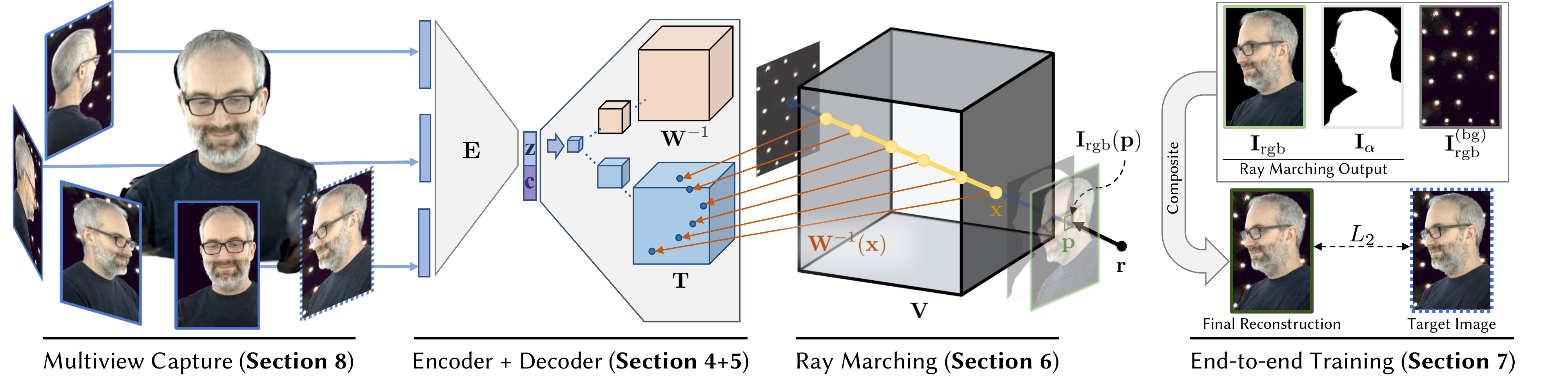}
    \end{center}
\caption{Pipeline of our method. We begin with a multi-view capture system, from which we choose a subset of the cameras as input to our encoder. The encoder produces a latent code $\mathbf{z}$ which is decoded into a volume that gives an RGB and $\alpha$ value for each point in space, as well as a warp field used to index into the RGB$\alpha$ volume. The decoder may optionally use an additional control signal (e.g., head pose) $\mathbf{c}$. We then use an accumulative ray marching algorithm to render the volume. The final output in image space is an RGB image with associated alpha mask and estimated background. We composite these components together and minimize the $L_2$ loss between the rendered and target images w.r.t.~network parameters.}
\label{fig:pipeline}
\end{figure*}

\section{Related Work} \label{sec:related_work}

Our approach is driven by learning and rendering techniques spanning multiple domains, from volumetric reconstruction and deformable volumes to neural rendering and novel view synthesis. The following paragraphs discuss similarities and differences to previous works in these areas.

\subsection{Classical Surface and Volumetric Reconstruction}


Point- and surface-based reconstruction techniques have a long history in computer vision (see~\cite{Furukawa:2015} for a review). Following successful efforts in stereo matching~\cite{Scharstein:2002}, most of the subsequent literature has focused on the multi-view case, including extensions of photometric consistency~\cite{Furukawa:2010} and depth map fusion~\cite{Merrell:2007,Zach:2007}. Despite some attempts at handling more complex materials or semi-transparent surfaces \cite{Szeliski:1999,Fitzgibbon:2005}, many popular multi-view stereo (MVS) methods, such as COLMAP \cite{Schoenberger:2016sfm,Schoenberger:2016mvs}, still struggle with thin structures and dense semi-transparent materials (e.g., hair and smoke). 

Volumetric reconstruction methods side-step this problem of explicit correspondence matching, beginning with Voxel Coloring~\cite{Seitz:1997,Prock:1998,Seitz:1999} and Space Carving~\cite{Kutulakos:2000,Broadhurst:2001,DeBonet:1999}. These methods recover occupancy and color in a voxel grid from multi-view images by evaluating the photo-consistency of each voxel in a particular order. Our method is similar to these classical voxel-based techniques in spirit, but rather than using a strict photometric consistency criterion, we learn a generative model that tries to best match the input images. Since we do not assume that objects in a scene are composed of flat surfaces, this approach also allows us to overcome the typical limitations of MVS methods and capture rich materials and fine geometry.

\subsection{Volumetric Reconstruction with Ray Potentials}

A number of more recent works on volumetric reconstruction have explored the concept of \emph{ray \mbox{potentials}}, i.e., cost functions between the first surface struck by a ray and the color (or other property) of the corresponding pixel.
\citet{Ulusoy:2015}, \citet{Savinov:2016}, and \citet{Paschalidou:2018} formulate graph-based energy or inference objectives using ray potentials as constraints. A differentiable ray consistency criterion that inspired our work is developed by~\citet{Tulsiani:2017,Tulsiani:2018}, who use an encoder-decoder architecture to predict voxel occupancy probabilities from RGB images. A loss on ray potentials evaluated in this volume is then backpropagated to the underlying convolutional architecture.

A fundamental difference between our work and previous works using ray potentials is that we model voxel transparency rather than occupancy probability, as we are focused on rendering rather than reconstruction. This explicit image formation process allows us to reconstruct and render dynamic scenes of semi-transparent materials, such as smoke.

\subsection{Deformable Volumes}
Non-rigidly deforming objects pose challenges to both optimization- and learning-based approaches. Over the years, significant efforts have been spent on their reconstruction and tracking from \mbox{RGB-D} sensors: the DynamicFusion method of~\citet{Newcombe:2015} produces a base 3D template surface using a Truncated Signed Distance Function (TSDF) representation, and a time-dependent warp to fuse a sequence of depth frames. \citet{Zollhofer:2014} and \citet{Innmann:2016} also deform a 3D template surface with an as-rigid-as-possible regularizer.

Our rendering model is based on classical volume ray marching~\cite{Levoy:1988,Kniss:2004}, but we introduce the concept of a \emph{warp field} to it: instead of directly sampling color and opacity along the ray, we first sample a 3D warp field encoding the location (within a template voxel grid) from which color and opacity are sampled. In 2D, similar techniques have been used to learn warps that align images~\cite{Jaderberg:2015,Dai:2017,Shu:2018}. In our work, the warp field is not only used to model motion but also increases the effective resolution of the voxel grid by simulating a dynamic non-uniform sampling grid.

\subsection{Neural Rendering}

Deep learning-based rendering has become an active area of exploration, with some methods relying on volumetric representations. \citet{Nguyen-Phuoc:2018} use a convolutional neural network applied on a voxel grid to produce an image, but their method requires correspondence between the images and the voxel reconstruction. DeepVoxels~\cite{Sitzmann:2018} automatically learns a 3D feature representation for novel view synthesis but is limited to static objects and scenes, and the employed architecture does not lend itself to real-time inference. \citet{Martin-Brualla:2018} use neural networks to fill holes and generally improve the quality of a textured geometric representation. \citet{Kim:2018} use a U-net architecture to convert an image of rasterized attributes to realistic images, similar to pix-to-pix~\cite{pix2pix2017}. Our method shares many similarities with these approaches but has one important difference: machine learning is only used to generate an RGBA volume, which is then rendered with a ray marching algorithm with no learned parameters. This is important as it gives us an interpretable volume, which may lead to better viewpoint generalization.

In our evaluation of hybrid rendering approaches, we employ the Deep Appearance Model of~\citet{Lombardi:2018} that provides a textured mesh representation. The method learns a Variational Autoencoder (VAE) model representing the dynamic mesh and view-dependent texture of a specific person's face. While we also evaluate our model on human faces, our method does not require precise mesh tracking or other forms of pre-processing. Instead, it is trained end-to-end, using only raw images as supervision. Volumetric representations such as ours are also able to better represent complex surfaces like hair that are difficult to model using meshes.

\subsection{Novel View Synthesis}

Novel view synthesis aims to produce RGB images of novel views given a set of input RGB images. Typically, these methods use a geometric proxy to assist in reprojecting 3D points back into the input images and some blending is performed to produce a final pixel color. \citet{Buehler:2001} and \citet{Davis:2012} use heuristics to blend contributions of different images based on the rays from the geometric proxy to each camera.  
\citet{Hedman:2018} uses neural networks to determine blending weights, which can overcome inaccurate geometric proxies. \citet{Zhou:2018} skip the geometric proxy altogether and use a neural network to compute blending weights of each image projected along a set of planes.
\citet{Penner:2017} compute a soft volumetric representation using MVS depth maps to perform novel view synthesis.
Unlike most novel view synthesis techniques, our method operates on sequences and creates an animatable model.


To compute geometric proxies, a multi-view stereo method is often employed. While free-viewpoint video methods \cite{Collet:2015,Prada:2016} rely on a sophisticated combination of multi-view stereo techniques (including silhouettes and MVS) to reconstruct many kinds of objects, our method creates novel views of dynamic scenes with a single generative framework. In addition, our model's latent embedding of the scene allows us to generate novel animations more flexibly by producing new embedding sequences.

\section{Overview} \label{sec:overview}




We present an end-to-end pipeline for rendering images from novel views with only image supervision that leverages an internal 3D volumetric representation. There are two main parts to the method: an encoder-decoder network that converts input images into a 3D volume $\mathbf{V}(\mathbf{x})$, and a differentiable raymarching step that renders an image from the volume $\mathbf{V}$ given a set of camera parameters. The method can be thought of as an autoencoder whose final layer is a fixed-function (i.e., no free parameters) volume rendering operation.

Formally, we model a volume that maps 3D positions, $\mathbf{x}\in\mathds{R}^3$, to a local RGB color and differential opacity
at that point,
\begin{equation}
\mathbf{V} : \mathds{R}^3 \rightarrow \mathds{R}^4, \qquad
\mathbf{V}(\mathbf{x}) = \left(\mathbf{V}_{\mathrm{rgb}}(\mathbf{x}), \mathbf{V}_\alpha(\mathbf{x})\right),
\end{equation}
where $\mathbf{V}_{\textrm{rgb}}(\mathbf{x})\in\mathds{R}^3$ is the color at $\mathbf{x}$ and $\mathbf{V}_\alpha(\mathbf{x})\in\mathds{R}$ is its differential opacity in the range $[0,\infty]$, with $0$ representing full transparency. The purpose of a semi-transparent volume is two-fold: first, it is a softening of a discrete volume representation which enables gradients to flow for learning; second, it allows us to model semi-transparent objects or bundles of thin structures that appear translucent at limited resolutions, like hair.

Fig.~\ref{fig:pipeline} shows a visual representation of the pipeline. We first capture a set of synchronized and calibrated video streams of a performance from different viewpoints. Next, an encoder network takes images from a subset of the cameras for each time instant and produces a latent code $\mathbf{z}$ that represents the state of the scene at that time. A volume decoder network produces a 3D volume given this latent code, $\mathbf{V}(\mathbf{x}; \mathbf{z})$, which yields an $\mathrm{RGB}\alpha$ value at each point $\mathbf{x}$. Finally, an accumulative ray marching algorithm renders the volume from a particular point-of-view. We train this system end-to-end by reconstructing each of the input images and minimizing the squared pixel reconstruction loss over the entire training set. At training time, we run through the entire pipeline to train the weights of the encoder-decoder network. At inference time, we produce a stream of latent codes $\mathbf{z}$ (either the sequence of latent codes produced by the training images or a novel generated sequence) and decode and render in real time.




\section{Encoder Network}\label{sec:encoder}

\begin{figure}[t]
    \begin{center}
    \includegraphics[width=1.0\columnwidth]{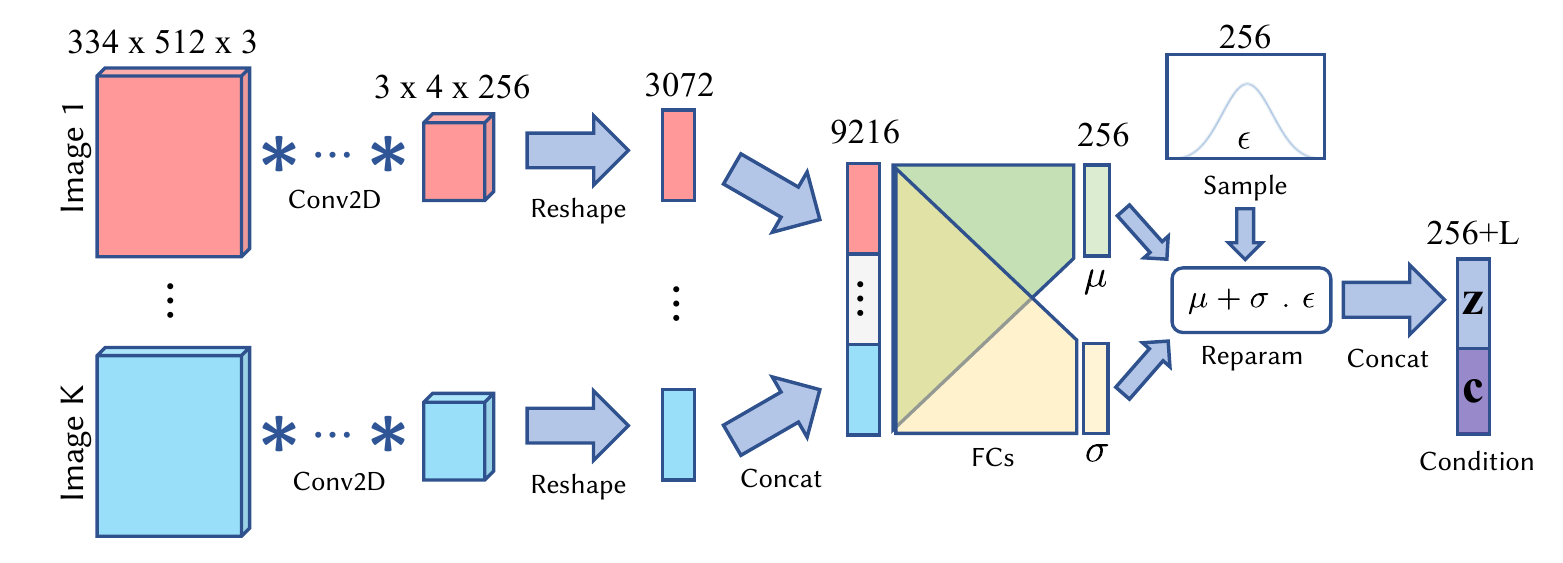}
    \end{center}
\caption{Encoder architecture. Inputs are images from a subset of K-cameras, passed through camera-specific CNNs. The latent variable $\mathbf{z}$ is concatenated with the L-dimensional conditioning variable $\mathbf{c}$ before being passed to the decoder (see \S\ref{sect:volume_decoders}). }
\label{fig:encoder}
\end{figure}

The main component of our system that enables novel sequence generation is the encoder-decoder architecture, where the scene's state is encoded using a consistent latent representation $\mathbf{z} \in \mathds{R}^{256}$. A traversal in this latent space can be decoded into a novel sequence of volumes that can then be rendered from any viewpoint (see \S\ref{sect:volume_decoders} for details). This is in contrast to methods that rely on specialized mesh constructions per frame which only allow for playback~\cite{Collet:2015} or limited control over the generative process~\cite{Prada:2016}. Moreover, this representation allows for conditional decoding, where only part of the scene's state is modified on playback (i.e., expression during speech, view-dependent appearance effects, etc.). The encoder-decoder architecture naturally supports this capability without requiring specialized treatment on the decoder side so long as paired samples of the conditioning variable are available during training.

To build the latent space, the information state of the scene at any given time is codified by encoding a subset of views from the multi-camera capture system using a convolutional neural network (CNN). The architecture of the encoder is shown in Fig.~\ref{fig:encoder}. Each camera view is passed through a dedicated branch before being concatenated with those from other views and further encoded down to the final shape. Although using all camera views as input is optimal from an information-theoretic perspective, we found that using $\text{K}=3$ views worked well in practice, while being much more memory and computationally efficient. To maximize coverage, we select a subset of cameras that are roughly orthogonal, although our system is not especially sensitive to the specific choice of views. In practice, we used the frontal, left and right-most camera views and downsampled the images by a factor of $8$ to size $334 \times 512$ pixels.   

To generate plausible samples during a traversal through the latent space, the generative model needs to generalize well between training samples. This is typically achieved by learning a smooth latent space. To encourage smoothness, we use a variational architecture~\cite{Kingma:CoRR:2013}. The encoder outputs parameters of a diagonal 256-dimensional Gaussian (i.e.,  $\mathbf{\mu}$ and $\mathbf{\sigma}$), whose KL-divergence from a standard Normal distribution is used as regularization. Generating an instance involves sampling from this distribution using the reparameterization trick, and decoding into the volumetric components described in \S\ref{sect:volume_decoders}.

In addition to encouraging latent space smoothness, the variational architecture also ensures that the decoder makes use of the conditioning variable when it is trained jointly with the encoder, as described in \S\ref{sec:training}. Specifically, since the variational bottleneck maximizes the non-informative latent dimensions~\cite{Higgins:2017}, information pertaining to the conditioning variable is projected out of the latent space, leaving the decoder no choice but to use the conditioning variable in its reconstruction. 


This method of conditioning can be applied using any auxiliary information available to the user for controlling the rendered output. In \S\ref{sec:Experiments} we show experiments demonstrating this for a few types of conditioning information. Of particular importance is view-conditioning, which allows view-dependent effects, such as specularity, to be rendered correctly. When viewed in VR, the auxiliary information in the form of a view-vector can be obtained from the relative orientation of the headset in the virtual scene.

\begin{figure}[t]
    \begin{center}
    \includegraphics[width=1.0\columnwidth]{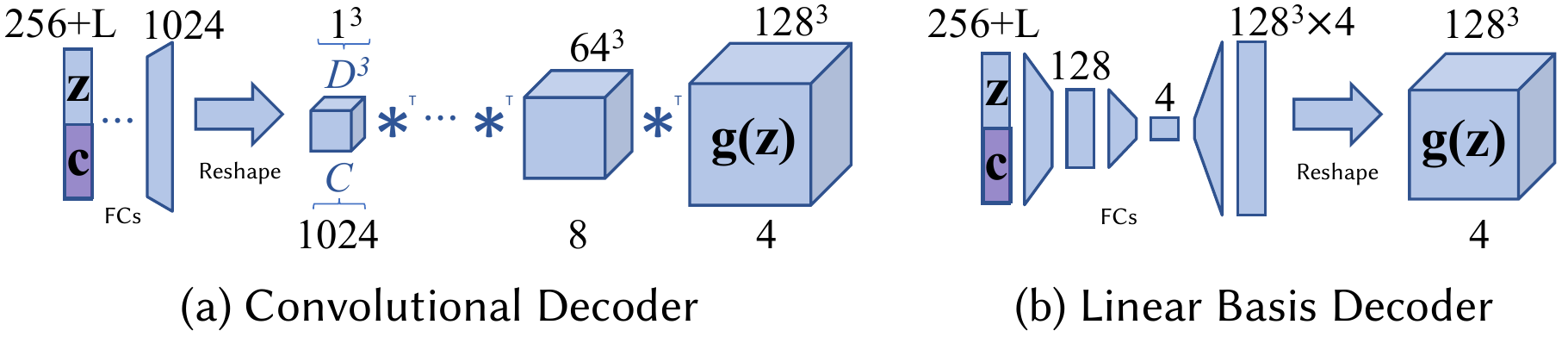}
    \end{center}
\caption{Voxel grid decoders. (a) Convolutional and (b) Linear basis decoders. $L$ denotes the size of the conditioning variable. }
\label{fig:voxelgrid_decoders}
\end{figure}

\section{Volume Decoders}
\label{sect:volume_decoders}
In this section we discuss parameterizing the $\mathrm{RGB}\alpha$ volume function~$\mathbf{V}$ using different neural network architectures which we call {\em volume decoders}. We discuss possible representations for the volumes (voxel grids and multi-layer perceptrons) and a method for increasing the effective resolution by using warping fields. Finally, we discuss view-conditioning for modeling view-dependent appearance.

\subsection{MLP Decoders}
One possible model for the volume function $\mathbf{V}(\mathbf{x}; \mathbf{z})$ at point $\mathbf{x}$ with state $\mathbf{z}$, is an implicit one with a series of fully connected layers with non-linearities. A benefit of this approach is that we're not restricted by voxel grid resolution or storage space. Unfortunately, in practice an MLP requires prohibitive size to produce high-quality reconstructions. We must also evaluate the MLP at every step along each ray in the ray-marching process (see \S\ref{sect:rendering}), imposing an equally restrictive upper bound on the MLP complexity for real-time applications.

\subsection{Voxel Grid Decoders}
Rather than trying to model the entire volume implicitly with an MLP, we may instead assume that the volume function can be modeled as a discrete 3D grid of voxels. We produce this explicit 3D voxel grid as the output tensor of a neural network. Let the tensor $\mathbf{Y}\in\mathds{R}^{C\times D \times D \times D}$ represent a $D{\times} D {\times} D$ grid of values in $\mathds{R}^C$, with $C$ the {\em channel} dimensions.
Define $S(\mathbf{x};\,\mathbf{Y}) : \mathds{R}^3 \rightarrow \mathds{R}^C$ to be an interpolation function that samples from the grid $\mathbf{Y}$ by scaling continuous values in the range $[{-}1,1]$ to the grid $[1,D]$ along each dimension, followed by trilinear interpolation. 
We can define a volume decoder for a 3D cube with center at $\mathbf{x}_o$ and sides of size $W$,
\begin{align}
\mathbf{V}(\mathbf{x}; \mathbf{z}) &= S\left(\frac{\mathbf{x}-\mathbf{x}_o}{W/2};\,\mathbf{g}(\mathbf{z})\right),
\end{align}
where $\mathbf{g}(\mathbf{z})$ is a neural network that produces a tensor of size $4{\times} D {\times} D {\times} D$. Note that we only evaluate the decoder function inside the volume it covers by computing intersections with its bounding volume. 

In practice, we use either a convolutional architecture or a series of fully-connected layers to implement ${\bf g}({\bf z})$. In the former case, we first apply a fully-connected layer and non-linearity to transform $\bf z$ into a $1024$-dimensional representation and reinterpret the resulting vector as a $1\times 1\times 1$ cube with $1024$ channels. 
We experiment with two convolutional architectures, one with a final size of $32^3$ (achieved with $5$ transposed convolutions and running at 90Hz) and one with a final size of $128^3$ (achieved with $7$ transposed convolutions and running at 22Hz).
As an alternative, we consider a bottleneck architecture capable running at 90Hz consisting of $3$ fully-connected layers with output sizes of $128$, $4$, and $128^3\times 4$, respectively, with the last layer representing the decoded volume. After decoding the volume, we apply a softplus function to the $\mathrm{RGB}\alpha$ values to ensure they are non-negative. Both decoder variants are illustrated in Fig.~\ref{fig:voxelgrid_decoders}.










\subsection{Warping Fields}

On their own, voxel grids are limited because they can only represent details as small as a single voxel and are computationally expensive to evaluate and store at high resolution. Additionally, they are wasteful in typical scenes where much of the scene consists of empty space. Common solutions to these problems are spatial acceleration structures like octrees (e.g., \citet{Riegler:2017}), but it's difficult to modify these to work in a learning setting. They also typically require that the distribution of objects within the structure is known \emph{a priori}, which is not true in our case as we do not use any 3D training data.

To solve these problems, we propose to use warping fields to both alter the effective resolution of the voxel volumes as well as model motion more naturally. In our warping formulation, we produce a template RGB$\alpha$ volume $\mathbf{T}(\mathbf{x})$ and a warp volume $\mathbf{W}^{-1}(\mathbf{x})$. Each point in the warp volume gives a corresponding location in the template volume from which to sample, making this an inverse warp as it maps from output positions to template sample locations. As before, both template and warp are decoded from a dynamic (per-frame) latent code $\mathbf{z}$, which we drop from the notation for conciseness.

The choice of inverse warps rather than forward warps allows representing resolution-increasing transformations by mapping a small area of voxels in the output space to a larger area in the template space without requiring additional memory. Thus, the inverse warp can represent details in the output space with higher resolution than uniform grid sampling, but remains well-defined everywhere in the output space, which is necessary for providing usable gradients during learning. 

Formally, we define the inverse warp field,
\begin{equation}
\mathbf{W}^{-1}(\mathbf{x}) \rightarrow \mathbf{y}\qquad\mathbf{x},\mathbf{y}\in\mathds{R}^3,
\end{equation}
where $\mathbf{x}$ is a 3D point in the output (rendering) space and $\mathbf{y}$ a 3D point in the RGB$\alpha$ template space. To generate the final volume value, we first evaluate the value of the inverse warp and then sample the template volume $\mathbf{T}$ at the warped point,
\begin{equation}
\mathbf{V}_{\mathrm{RGB}\alpha}(\mathbf{x}) = \mathbf{T}_{\mathrm{RGB}\alpha}(\mathbf{W}^{-1}(\mathbf{x})).
\end{equation}

\subsection{Mixture of Affine Warps}

An important piece of including warp fields is determining how they're produced. As we show later, the architecture of the warp field decoder makes a large difference on the quality of the model.

A straightforward approach to decoding warp fields would be to use deconvolutions to produce a warp field with freely-varying template sample points at each output point. This parameterization, however, is too flexible, resulting in overfitting and poor generalization to novel views as we show in our experimental results. We instead take the approach that the basic building block of a warp field should be an affine warp. Since a single affine warp can't model non-linear bending, we use a spatial mixture of affine warps to produce an inverse warp field.

We write the affine mixture as,
\begin{equation}\label{eq:affinemix}
\mathbf{W}^{-1}(\mathbf{x}) = \sum_i \mathbf{A}_i\left(\mathbf{x}\right) a_i(\mathbf{x}),
\end{equation}
\begin{equation}\label{eq:affinemix2}
\textrm{with}\quad \mathbf{A}_i(\mathbf{x}) = \mathbf{R}_i \left( \mathbf{s}_i \circ \left(\mathbf{x} - \mathbf{t}_i\right) \right),\qquad
a_i(\mathbf{x}) = \frac{w_i \left(\mathbf{A}_i\left(\mathbf{x}\right)\right)}{\sum_j w_j \left(\mathbf{A}_j\left(\mathbf{x}\right)\right)},
\end{equation}
where $\mathbf{A}_i(\mathbf{x})$ is the $i^{\mathrm{th}}$ affine transformation, $\{\mathbf{R}_i, \mathbf{s}_i, \mathbf{t}_i\}$ define the rotation, scaling, and translation of the $i^{\mathrm{th}}$ affine transformation parameters, $\circ$ is element-wise multiplication, and $w_i(\mathbf{x})$ is the weight volume of the $i^{\mathrm{th}}$ warp. Note that we sample the spatial mixture weight $w_i$ after warping (``warped weights'').
The intuition behind this is that the warped space represents different parts of the scene, and the weighting function should be in that space as well. This can be viewed as an extension of linear blend skinning~\cite{Lewis:2000} applied to a volumetric space.

To compute the transformation parameters $\{\mathbf{R}_i, \mathbf{s}_i, \mathbf{t}_i\}$ and the weighting volume $w_i$ we use 2 fully-connected layers after the encoding $\mathbf{z}$. For rotation, we produce a rotation quaternion vector which is normalized and transformed into a rotation matrix. Before outputting the values of the weighting volume, we apply $\exp(\cdot)$ to ensure the weights are non-negative. Unlike the voxel volume $\mathbf{V}$, we clamp samples outside the weighting volume $w_i$ to the surface otherwise the warps can get ``stuck'' early in training if they land outside the volume. In practice, we found that a mixture of 16 warps provides sufficient expressiveness. In all experiments where warping is used, we learn an additional global warp $\{\mathbf{R}_{\mathrm{g}}, \mathbf{s}_{\mathrm{g}}, \mathbf{t}_{\mathrm{g}}\}$ (with parameters produced by MLP) that is applied to $\mathbf{x}$ before the warping field, and we use a $32^3$ voxel grid to represent the warp field as we found that the low resolution provides smoothness from the trilinear interpolation that helps learning.

\subsection{View Conditioning}

In order to model view-dependent appearance, we opt to condition the $\mathrm{RGB}$ decoder network on the viewpoint. This allows us to model specularities in a data-driven way, without specifying a particular functional form. To do this, we input the normalized direction of the camera to the decoder alongside the encoding. Note that for view-conditioned models, we use separate convolutional branches to produce the $\mathrm{RGB}$ values and $\alpha$ values as we only want to condition the $\mathrm{RGB}$ values on the viewpoint.

\begin{algorithm}[t]
\SetAlgoNoLine
\SetEndCharOfAlgoLine{}
${\mathbf{I}}_{\textrm{rgb}}$=$\mathbf{0}$; $\,$  ${\bf I}_\alpha$=0; $\,$  $t$=$t_\mathrm{min}$;\;
\Repeat{${\bf I}_\alpha {=} 1$ or $t{>}t_\mathrm{max}$}{
${\mathrm{d}\alpha(t)} = \min\left({\bf I}_\alpha +  \Delta\mathbf{V}_{\alpha}(\mathbf{r}_o + t \mathbf{r}_d), 1\right) - {\bf I}_\alpha$\;
${\mathbf{I}}_{\textrm{rgb}} = {\mathbf{I}}_{\textrm{rgb}} +\mathbf{V}_{\mathrm{rgb}}(\mathbf{r}_o + t \mathbf{r}_d)\, \mathrm{d}\alpha(t)$\;
${\bf I}_\alpha = {\bf I}_\alpha + {\mathrm{d}\alpha(t)}$\;
$t = t + \Delta$\;
}
\Return ${\bf I}_{\textrm{rgb}}, {\bf I}_\alpha$
\caption[Ray Accumulation]{Accumulative rendering of ray $\mathbf{r}_o{+}t\mathbf{r}_d$ with segment $t{\in}[t_\mathrm{min}, t_\mathrm{max}]$ intersecting~$\mathbf{V}$. (Integration of Eq.~\eqref{eq:integral_color} with stepsize $\Delta$.)}
\label{alg:imageformation}
\end{algorithm}


\section{Accumulative Ray Marching}
\label{sect:rendering}
We formulate a rendering process for semi-transparent volumes that mimics front-to-back additive blending:
As a camera ray traverses a volume of inhomogeneous material, it accumulates color in proportion to the local color and density of the material at each point along its path. 
\subsection{Semi-Transparent Volume Rendering}

We generate images from a volume function $\mathbf{V}(\mathbf{x})$ by marching rays through the volume it models. To model occlusions, the ray accumulates not only color but also opacity. If the accumulated opacity reaches~$1$ (for example, when the ray traverses an opaque region), then no further color can be accumulated on the ray. In particular, the color ${\mathbf{I}}_{\textrm{rgb}}(\mathbf{p})$ at a pixel $\mathbf{p}$ in the focal plane of a camera with center~$\mathbf{r}_o\in\mathds{R}^3$ and image-to-world transformation $\mathbf{P}^{-1}$ is given by raymarching in the unit direction~$\mathbf{r}_d=(\mathbf{P}^{-1}\mathbf{p}-\mathbf{r}_o)/\lVert \mathbf{P}^{-1}\mathbf{p}-\mathbf{r}_o\rVert\in\mathds{R}^3$. This leads to the rendering process
\begin{align}
{\mathbf{I}}_{\textrm{rgb}}(\mathbf{p}) &= \int_{t_\mathrm{min}}^{t_\mathrm{max}} \mathbf{V}_{\mathrm{rgb}}\left(\mathbf{r}_o + t\mathbf{r}_d\right) \frac{\mathrm{d}\alpha(t)}{\mathrm{d}t}\ \mathrm{d}t, \label{eq:integral_color} \\
\textrm{with}\quad \alpha(t) &= \min\left(\int_{t_\mathrm{min}}^{t} \mathbf{V}_{\alpha}\left(\mathbf{r}_o + s \mathbf{r}_d\right) \ \mathrm{d}s,1\right) \label{eq:integral_alpha},
\end{align}
where $t{\in}[t_\mathrm{min},t_\mathrm{max}]$ denotes the segment of the ray that intersects the volume modeled by $\mathbf{V}$, and $\min(\cdot,1)$ in Eq.~\eqref{eq:integral_alpha} ensures that the accumulated ray opacity is clamped at $1$. Furthermore, we set the final image opacities to ${\bf I}_\alpha({\bf p}) = \alpha(t_{\max})$.

Algorithm~\ref{alg:imageformation} shows the computation of the output color for a ray intersecting the volume $\mathbf{V}$, and represents the numerical integration of Eqs.~(\ref{eq:integral_color}--\ref{eq:integral_alpha}) using the rectangle rule\footnote{We use the rectangle rule twice with samples on the right of each step interval and backwards differences to discretize the derivative of $\alpha(t)$.}.
Importantly, this image formation model is differentiable, which allows us to optimize the parameters of the volume $\mathbf{V}$ to match target images.
In practice, we set the step size to be $\frac{1}{128}^{\mathrm{th}}$ the size of the volume, which provides adequate voxel coverage and allows the rendering process to run at 90Hz in an OpenGL shader at $512 \times 600$ resolution, enabling real-time stereo in Virtual Reality.

\subsection{Hybrid Rendering}
While the semi-transparent volume representation is very versatile, certain
kinds of scene content can be more efficiently represented at high resolution using surface-based representations combined with unwrapped texture maps. One example is fine detail in human faces, for which specialized capture systems are available and commonly use texture resolutions larger than $1024^2$~\cite{Beeler2011,Lombardi:2018,Fyffe2017}.

The volume representation described above offers a natural integration with
such existing mesh-based representations. Rendering proceeds as described in
Algorithm~\ref{alg:imageformation} with one modification: we set $t_{\mathrm{max}}$ to the mesh depth for all rays that intersect the mesh. Whenever one of these rays reaches $t_{\mathrm{max}}$, any remaining color throughput is filled with the color of the mesh at the intersection. 

This technique can also be used during learning to avoid expending representational power in these parts of the scene. As we show in \S\ref{subsec:meshvoxel}, the resulting semi-transparent volume learned using the hybrid rendering process described above naturally avoids occluding the mesh in areas where the mesh provides a higher-fidelity representation. 

\section{End-to-end Training}\label{sec:training}

In this section we discuss the details of training our method. Training the system consists of training the weights $\theta$ of the encoder-decoder network. We discuss the estimation of a per-camera color calibration matrix and static background image, the construction of our loss function, and reconstruction priors that improve accuracy.

\subsection{Color Calibration}

Although we have geometrically calibrated the cameras in our multi-view system, we have not color calibrated them relative to one-another. We need to ensure that one radiance value will be converted to the same pixel value for each of the cameras. To do this, we introduce a per-camera and per-channel gain $g$ and bias $b$ that is applied to our reconstructed image before comparing to ground truth. This allows our system to account for slight differences in overall intensity in the image.

\subsection{Backgrounds}

In our training data we often have static backgrounds that the algorithm will try to reconstruct. To ensure that our algorithm only reconstructs the object of interest, we estimate a per-camera background image $\mathbf{I}_{\mathrm{rgb}}^{(\mathrm{bg})}$. The background image is static across the entire sequence, capturing only stationary objects that are generally outside of the reconstruction volume. 

We obtain a final image $\widehat{\mathbf{I}}_{\mathrm{rgb}}$ from a specific view by raymarching all pixels ${\bf p}$ according to Eq.~\eqref{eq:integral_color} and merging ${\bf I}_{\mathrm{rgb}}({\bf p})$ with its corresponding background pixel $ \mathbf{I}_{\textrm{rgb}}^{(\mathrm{bg})}(\mathbf{p})$ according to the remaining opacity when exiting the volume,
\begin{equation}
\widehat{\mathbf{I}}_{\mathrm{rgb}}(\mathbf{p}) = \left(1 - \mathbf{I}_{\alpha}(\mathbf{p})\right) \mathbf{I}_{\textrm{rgb}}^{(\mathrm{bg})}(\mathbf{p})+\left(g {\bf I}_{\mathrm{rgb}}({\bf p}) + b\right).
\end{equation}
This background estimation process greatly reduces the amount of artifacts in the reconstruction.

\subsection{Reconstruction Priors}

\begin{figure}[t]
    \begin{center}
    \includegraphics[width=0.47\textwidth]{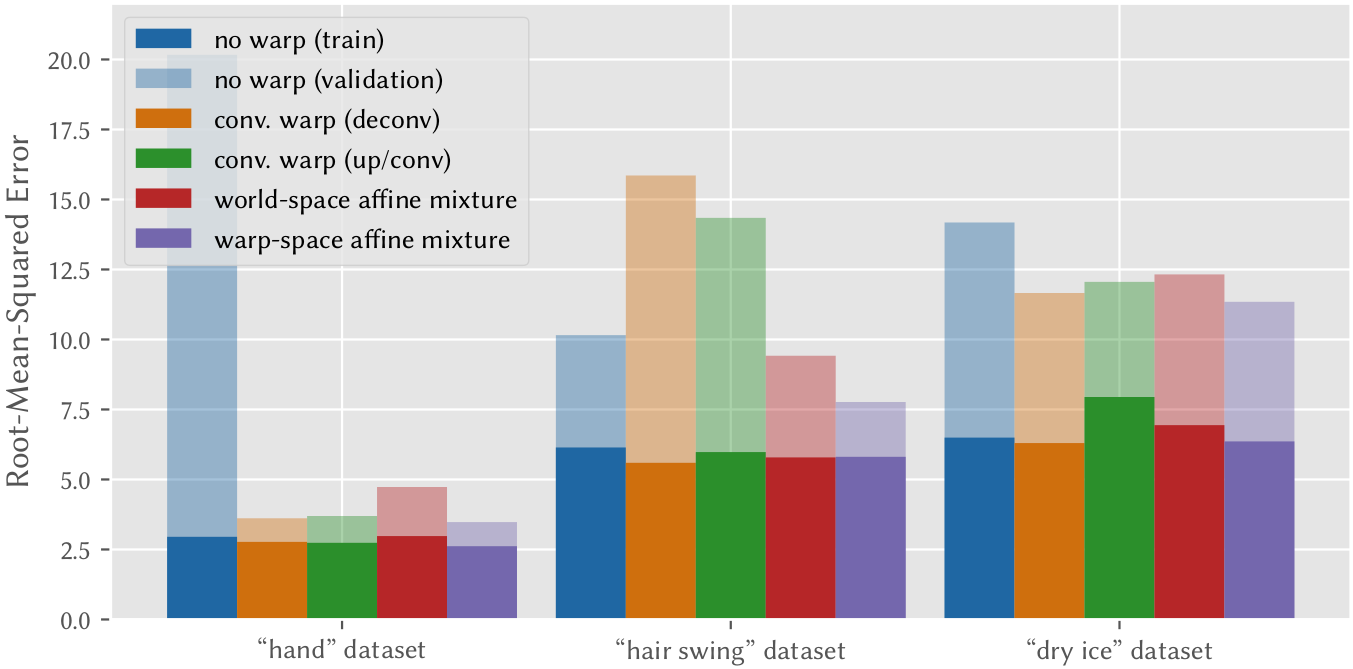}
    \end{center}
\caption{Warping method evaluation. In this experiment we evaluate different techniques for warping. We evaluate on three different datasets: a moving hand, swinging hair, and dry ice smoke. We evaluate models with: no warping, convolutional warp (using transposed convolutions), convolutional warp (using trilinear upsampling followed by convolution), no warp affine mixture, where the spatial weighting volume $w_i$ is not warped before evaluating (Eq.~\eqref{eq:affinemix_worldspace}), and warp-space affine mixture, as described in Eq.~\eqref{eq:affinemix2}. In all cases, our warp-space affine mixture outperforms the other models.}
\label{fig:warpeval}
\end{figure}

Without using priors, the reconstructed volumes tend to include smoke-like artifacts. These artifacts are caused by slight differences in appearance from different viewpoints due to calibration errors or view-dependent effects. The system can learn to compensate for these differences by adding a small amount of opacity that becomes visible from one particular camera. To reduce these artifacts, we introduce two priors. 

The first prior regularizes the total variation of the log voxel opacities,
\begin{equation}
J_{\textrm{tv}}({\bf V}_\alpha) = \frac{1}{N} \sum_{\mathbf{x}} \lambda_{\mathrm{tv}} \left\| \frac{\partial}{\partial \mathbf{x}} \log {\bf V}_{\alpha}(\mathbf{x}) \right\|,
\end{equation}
where the sum is performed over all the voxel centers $\mathbf{x}$ and $N$ is the number of voxels.
This term helps recover sharp boundaries between opaque and transparent regions by enforcing sparse spatial gradients.
We apply this prior in log space to increase the sensitivity of the prior to small $\alpha$ values because the artifacts tend to be mostly transparent.

The second prior is a beta distribution $\left(\textrm{Beta}(0.5,0.5)\right)$ on the final image opacities $\mathbf{I}_{\alpha}({\bf p})$. We write the regularization term using the negative log-likelihood of the beta distribution,
\begin{equation}
J_{\textrm{B}}\left({{\bf I}}_\alpha\right)=\frac{1}{P} \sum_{\mathbf{p}} \lambda_{\textrm{B}} \left[ \log\left(\mathbf{I}_{\alpha}(\mathbf{p})\right) + \log\left(1 - \mathbf{I}_{\alpha}(\mathbf{p})\right) \right],
\end{equation}
where $\mathbf{p}$ is an image pixel and $P$ is the number of pixels. This prior reduces the entropy of the exit opacities and is based on the intuition that most of our rays should strike the object or the background; fewer rays will graze the surface of the object, picking up some opacity but not saturating.

\subsection{Training Hyperparameters}

Our full training objective is
\begin{align}
\ell(\theta) = & \frac{1}{P} \sum_{\mathbf{p}} \left\| \widehat{{\bf I}}_{\mathrm{rgb}}(\mathbf{p}) - \mathbf{I}^{*}_\mathrm{rgb}(\mathbf{p}) \right\|^2 + \nonumber \\
& \lambda_{\mathrm{KL}} D_{\mathrm{KL}}\left(\mathbf{z}\,\|\,\mathcal{N}(0, 1) \right) + J_{\textrm{tv}}({\bf V}_\alpha) + J_{\textrm{B}}\left({\bf I}_\alpha\right),
\end{align}
where $\mathbf{I}^{*}_\mathrm{rgb}(\mathbf{p})$ is the ground truth image and $D_{\mathrm{KL}}(\mathbf{z}\,\|\,\mathcal{N}(0, 1))$ is the KL divergence between the latent encoding $\mathbf{z}$ and a standard normal distribution used in a variational autoencoder~\cite{Kingma:CoRR:2013}.

We use Adam~\cite{Kingma:2014} to optimize the loss function. In our experiments, we set $\lambda_{\mathrm{KL}} = 0.001$, $\lambda_{\mathrm{tv}} = 0.01$, and $\lambda_{\mathrm{B}} = 0.1$ and we use a batch size of $16$ with a fixed learning rate of $10^{-4}$ on the encoder-decoder network weights (the estimated background images $\mathbf{I}^{(\mathrm{bg})}$ and per-camera gain and bias $g, b$ are given separate learning rates of $10^{-1}$ and $10^{-3}$, respectively). We randomly sample $128 \times 128$ pixels from the image to reduce memory usage. While raymarching, we compute step sizes in normalized voxel space (i.e., $[-1, 1]^3$) as it makes the end of the ray the expected saturation point at network initialization. We train for about 500,000 iterations, depending on dataset size, which takes 10 days on a single NVIDIA Tesla V100.

\begin{table}[t]
\caption{View conditioning comparison. Conditioning the decoder on viewpoint improves the reconstruction on a set of validation viewpoints.}
\begin{tabular}{lll}
\toprule
                 \hspace{50mm}      & \multicolumn{2}{c}{Face} \\
\cmidrule{2-3}
                       & Train         & Val.          \\
\midrule
No view conditioning      & 51.1         & 117.8         \\
View conditioning   & \textbf{38.7}          & \textbf{85.7}         \\
\bottomrule
\end{tabular}
\label{table:viewconditioning}
\end{table}

\section{Experiments}\label{sec:Experiments}

We perform a number of quantitative and qualitative experiments to validate our model. For all experiments, we use the convolutional volume decoder with size $128^3$. We show that the design choices of our model provide a good compromise between quality and speed, and that the model generalizes to new viewpoints. We show results of our method on objects that are typically hard to reconstruct (e.g., fuzz, smoke, and hair) and demonstrate the method combined with traditional triangle rasterization. Finally, we demonstrate animation of our models by interpolating in the latent space and driving the reconstruction from user input.

To capture data, we used a multi-camera capture system consisting of 34 $4096\times2668$-resolution 30Hz color cameras placed on a hemisphere with a radius of approximately one meter. We calibrate the camera system using an icosahedral checkerboard pattern~\cite{Ha:2018}. The raw images and camera calibration are then used as the only input to our method.

In our quantitative experiments, we compare mean-squared error of pixel reconstructions on the training cameras and also on a set of 7 held-out validation cameras. This allows us to test how well our model extrapolates to novel viewpoints.

\begin{table}[t]
\caption{Evaluation of priors and background model. We show training and validation MSE for the ``fuzzy toy'' object with and without priors and using a known background, learned background, and no background model. Surprisingly, the learned background model with priors outperforms a known background.}
\begin{tabular}{lll}
\toprule
\hspace{5cm}           & \multicolumn{2}{c}{Fuzzy Toy} \\
\cmidrule{2-3}
Background Model                       & Train         & Val.          \\
\midrule
Known BG / priors      & 87.4         & 197.6         \\
Known BG / no priors   & \textbf{76.1}          & 330.5         \\
\midrule
Learned BG / priors    & 115.4          & \textbf{183.7}        \\
Learned BG / no priors & 94.1          & 281.7         \\
\midrule
No BG / priors         & 208.8         & 386.7         \\
No BG / no priors      & 86.6          & 727.6        \\
\bottomrule
\end{tabular}
\label{table:priorbg}
\end{table}

\subsection{Warping Method}

To validate our warp representation, we compare against several variants: a model with no warping, a model with a warp produced by a convolutional neural network, a model that does not apply the warp before computing the affine mixture weight, and the proposed affine mixture warp model (i.e., Eq.~\eqref{eq:affinemix}).

\begin{figure*}[t]
    \begin{center}
    \includegraphics[width=0.30\textwidth]{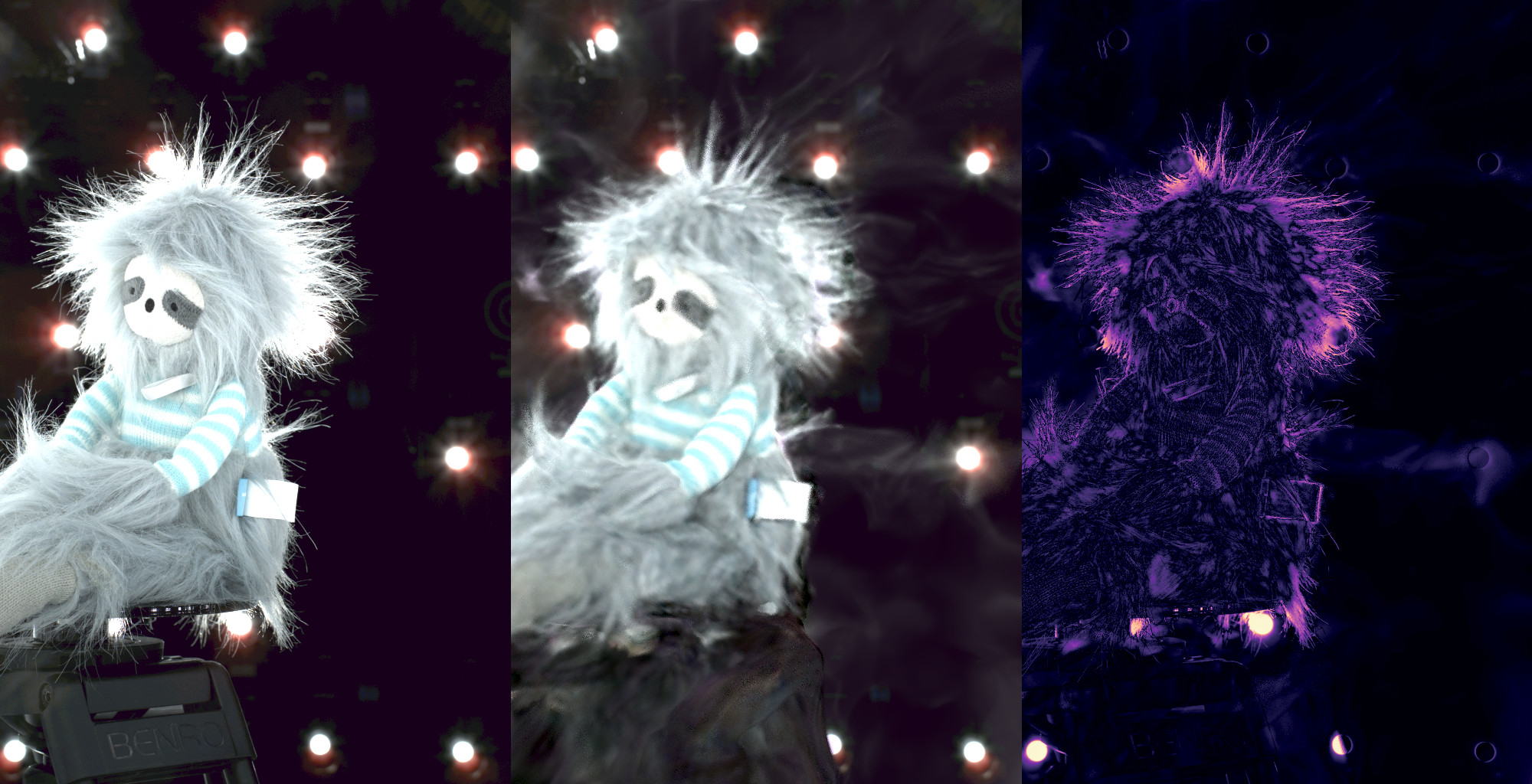}~
    \includegraphics[width=0.30\textwidth]{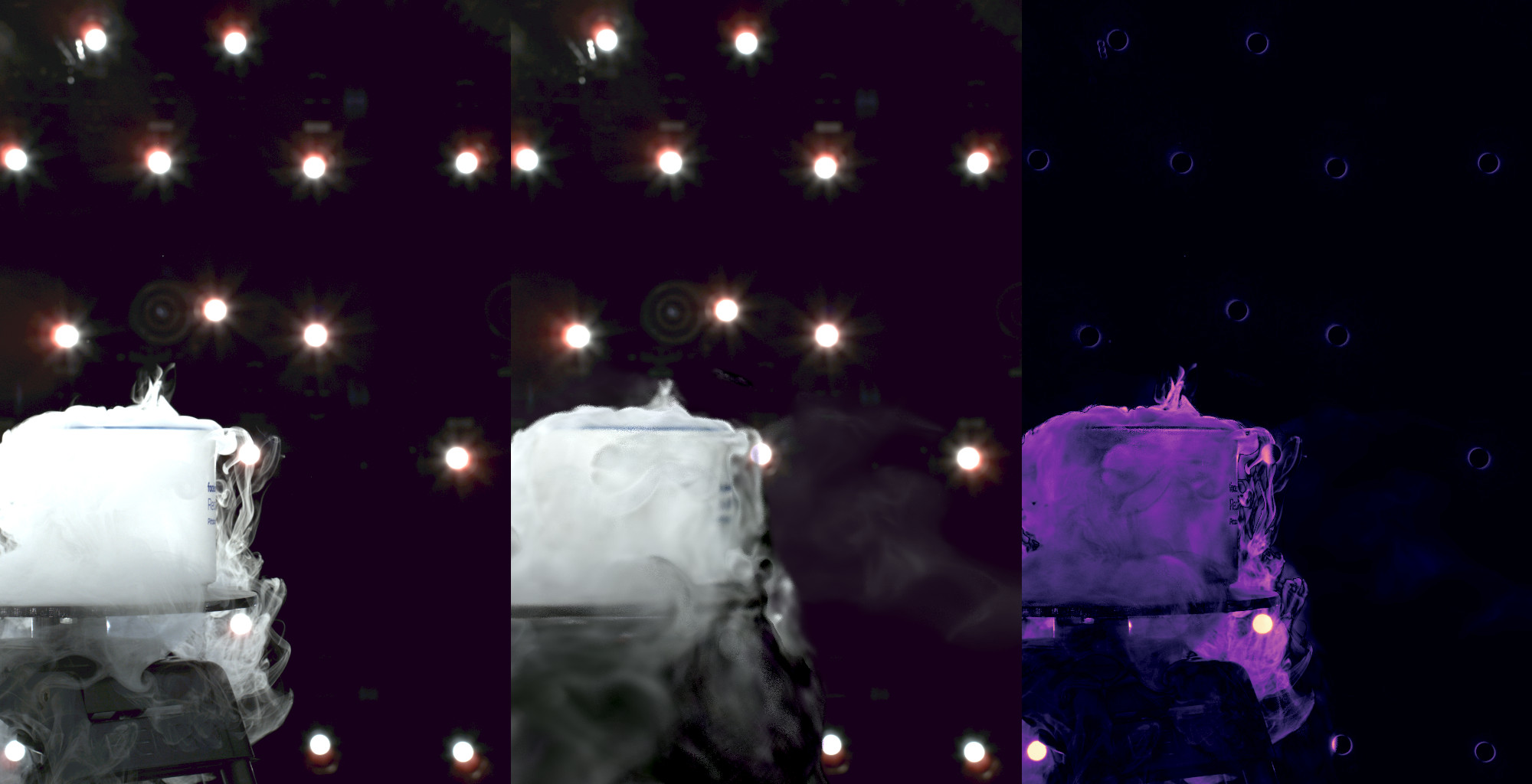}~
    \includegraphics[width=0.30\textwidth]{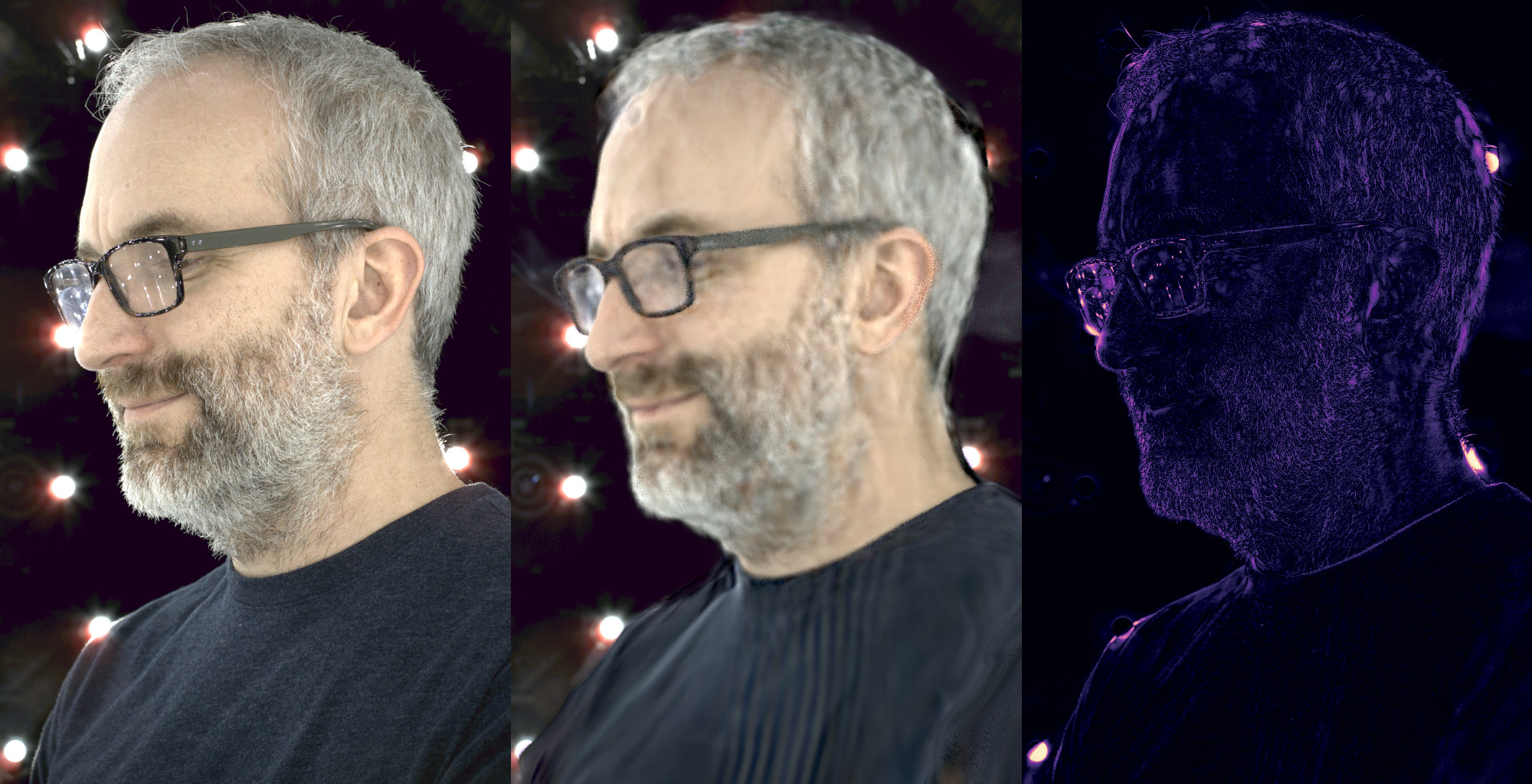}~
    \includegraphics[width=0.025\textwidth]{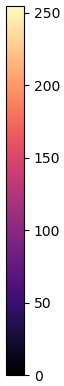}\\
    \includegraphics[width=0.30\textwidth]{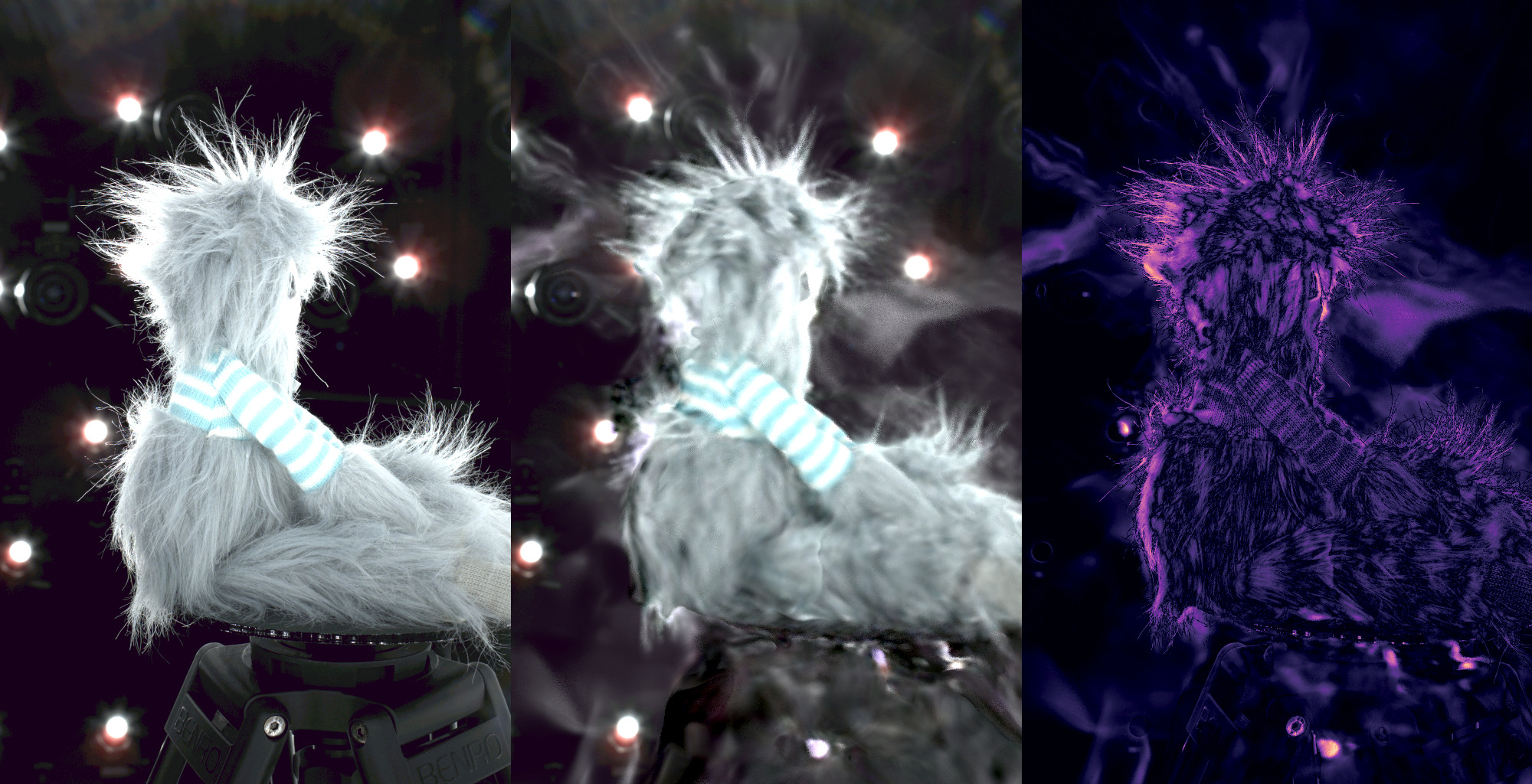}~
    \includegraphics[width=0.30\textwidth]{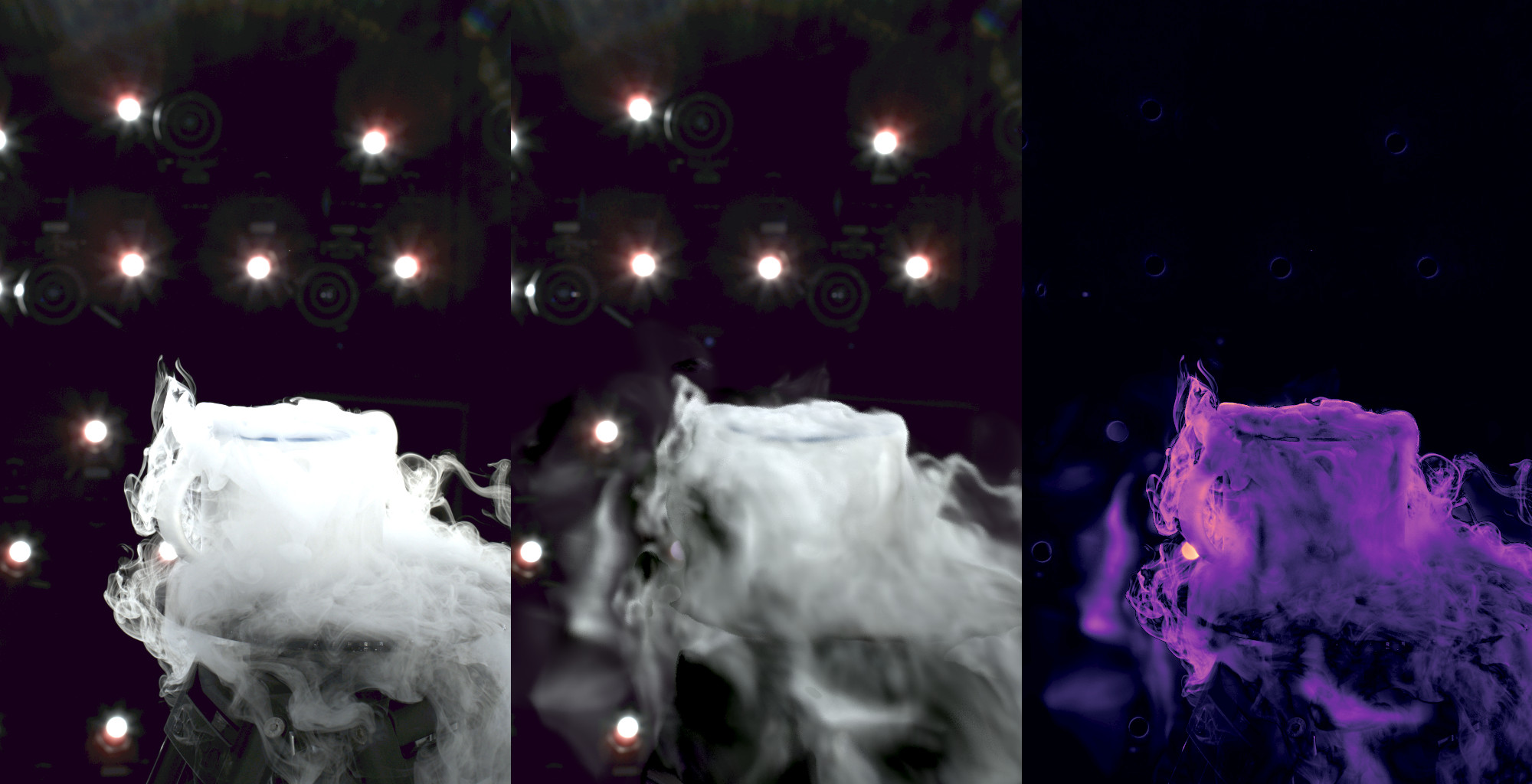}~
    \includegraphics[width=0.30\textwidth]{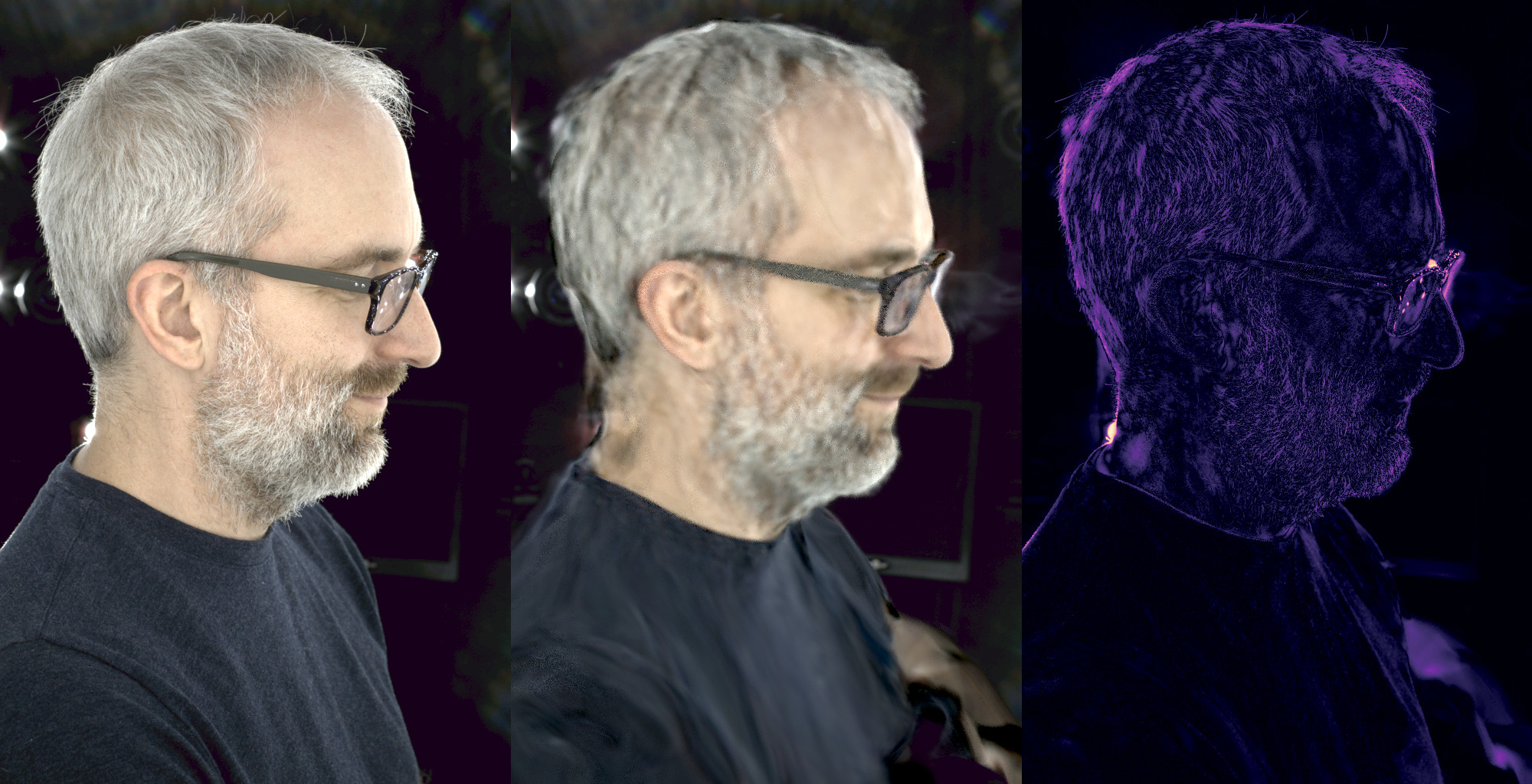}~
    \includegraphics[width=0.025\textwidth]{figs/qual/colorbar3.png}\\
    \includegraphics[width=0.30\textwidth]{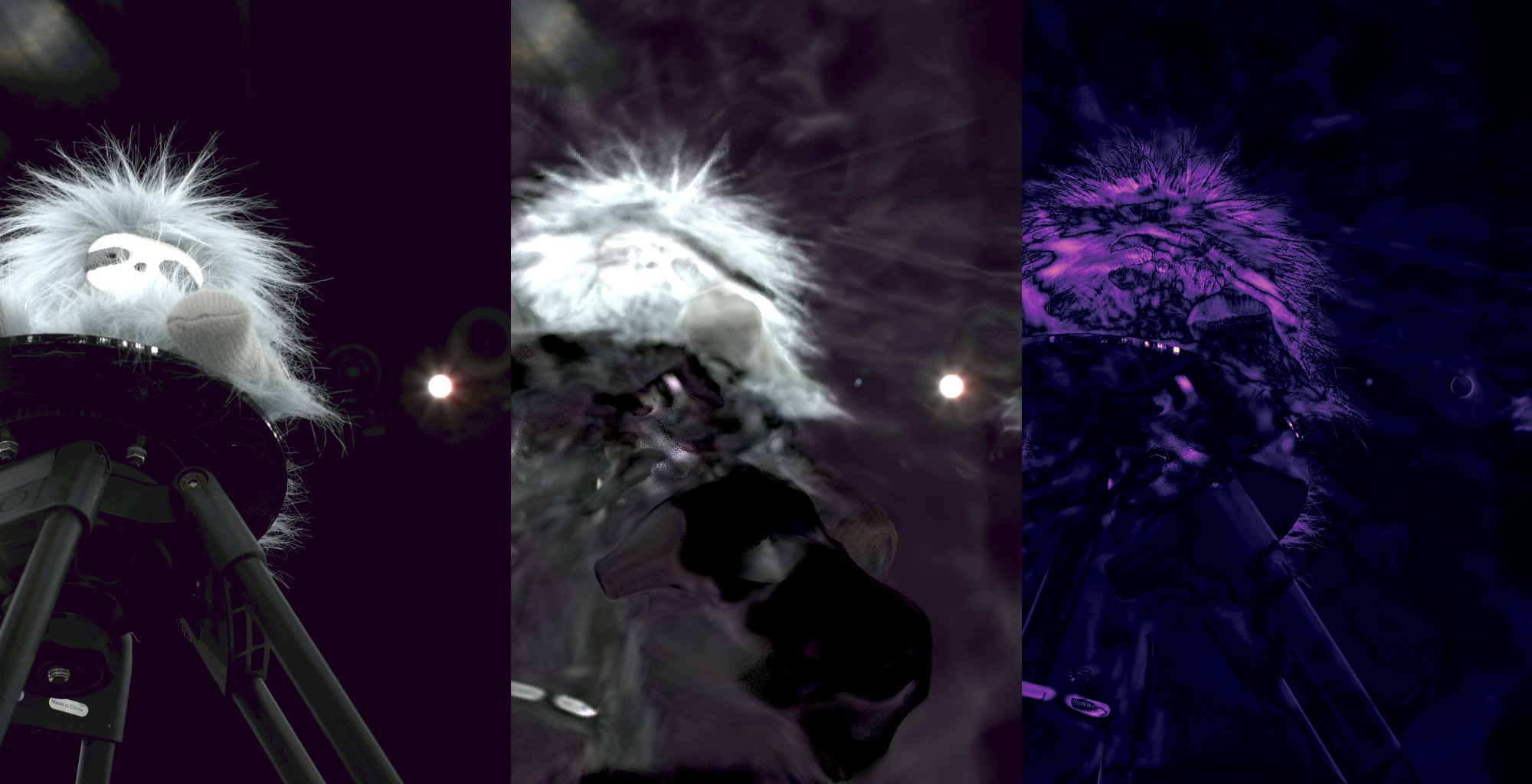}~
    \includegraphics[width=0.30\textwidth]{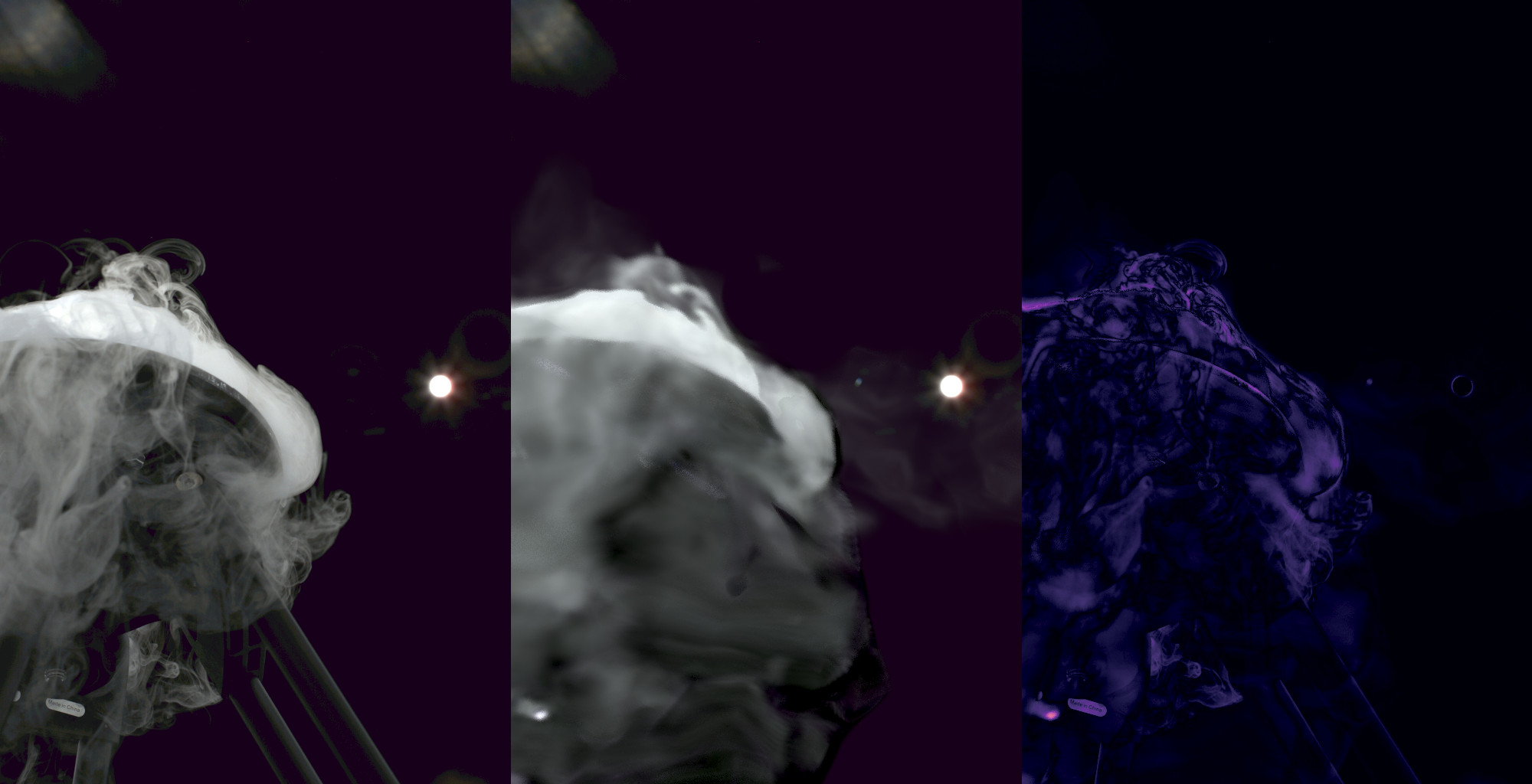}~
    \includegraphics[width=0.30\textwidth]{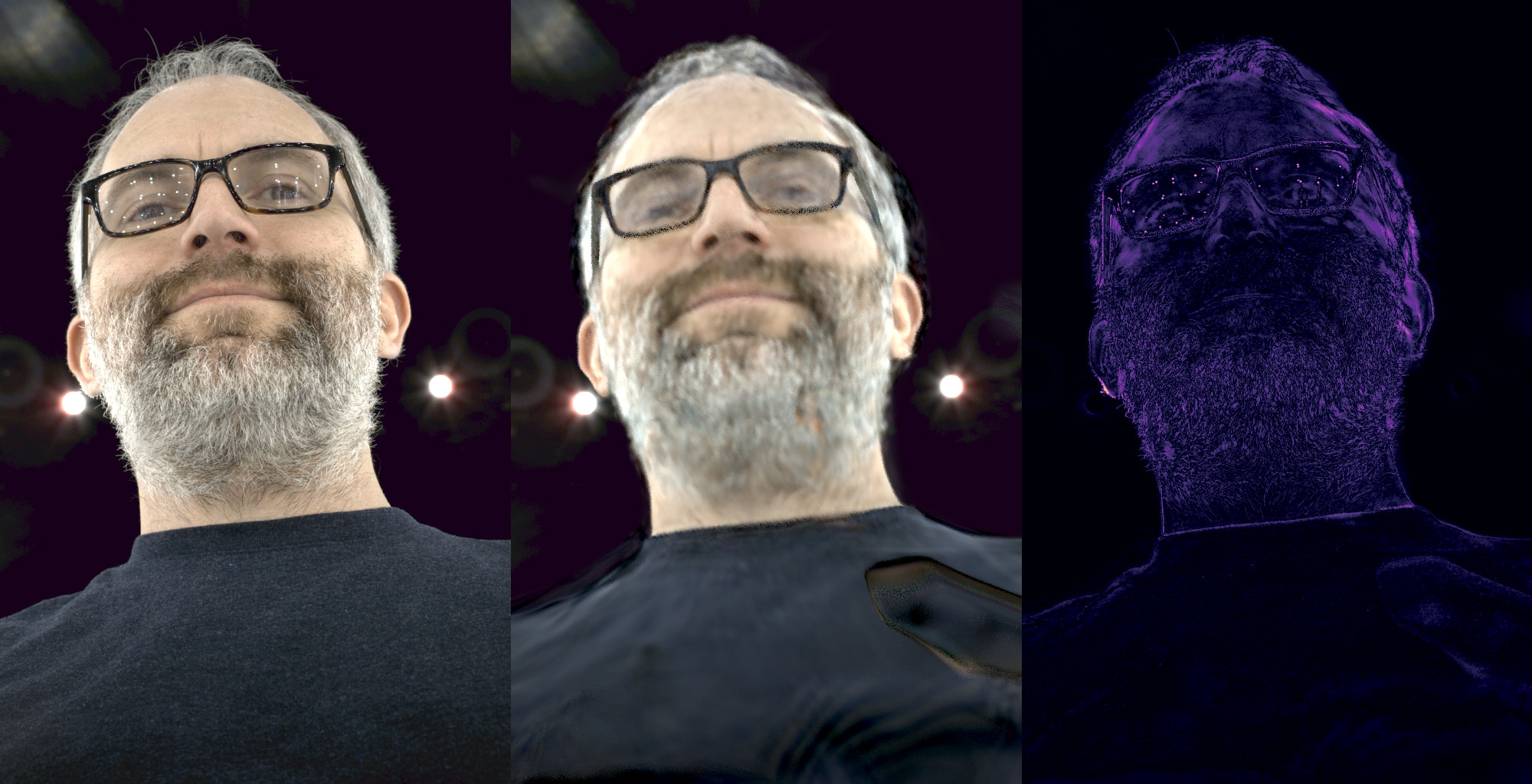}~
    \includegraphics[width=0.025\textwidth]{figs/qual/colorbar3.png}\\    
    \end{center}
\caption{Qualitative results. In this figure we show ground truth, reconstructed images, and a visualization of the root-mean-squared error for each pixel for 3 datasets on 3 held-out validation viewpoints.}
\label{fig:qualitative}
\end{figure*}

For the model that does not apply the warp to compute the mixture, we modify Eq.~\eqref{eq:affinemix2} as follows:
\begin{equation}\label{eq:affinemix_worldspace}
\textrm{with}\quad A_i(\mathbf{x}) = \mathbf{R}_i \left( \mathbf{s}_i \circ \left(\mathbf{x} - \mathbf{t}_i\right) \right),\qquad
a_i(\mathbf{x}) = \frac{w_i\left(\mathbf{x}\right)}{\sum_j w_j\left(\mathbf{x}\right)}.
\end{equation}
The main difference is that the mixture is done in ``world'' space rather than in warped space. Since the mixture weights are not warped, the decoder must match any motion of the template by moving the mixture weights.

Fig.~\ref{fig:warpeval} shows the results of our warping evaluation on three datasets: a moving hand, swinging hair, and dry ice smoke, each approximately 20 seconds long. In each dataset, our affine mixture model outperforms models with no warp and outperforms models with a convolutional warp field on the validation cameras. In particular, the convolutional warp fields completely fail on the ``hair swing'' dataset. The results also show that modeling the weighting volume in warp space is better than in ``world'' space.

\subsection{View-Conditioning}

An important part of rendering is being able to model view-dependent effects such as specularities. To do this, we can condition the $\mathrm{RGB}$ decoder $\mathbf{V}_{\mathrm{rgb}}$ on the viewpoint of the rendered view. This allows the network to change the color of certain parts of the scene depending on the angle it's viewed from.

Table \ref{table:viewconditioning} shows a quantitative experiment comparing a view-conditioned model to a non-view-conditioned model for a human face. The results show that the view-conditioned model is better able to model the scene from novel viewpoints. This happens not only because the view-conditioned model can represent some of the view-dependent appearance of the scene, but also because the non-view-conditioned model incorrectly reconstructs extra semi-transparent voxels near the surface of the object to explain view-dependent phenomena.

\subsection{Priors and Background Estimation}

We impose several priors on the reconstructed volume to help reduce the occurrence of artifacts in the reconstruction. In this experiment, we evaluate the effectiveness of background estimation and the priors on the reconstructed volume quality in terms of mean-squared-error on the validation viewpoints. We evaluate 3 different scenarios: known background image, learned background image, and no background model, each scenario evaluated with and without priors.

Table \ref{table:priorbg} shows the results of this experiment. For each background setting, using the priors improves performance on the validation viewpoints. Surprisingly, learning a background image outperformed using a known background image, but the improvement is small.
 
\subsection{Qualitative Results}

\begin{figure*}
    \centering
    \includegraphics[width=1.0\textwidth]{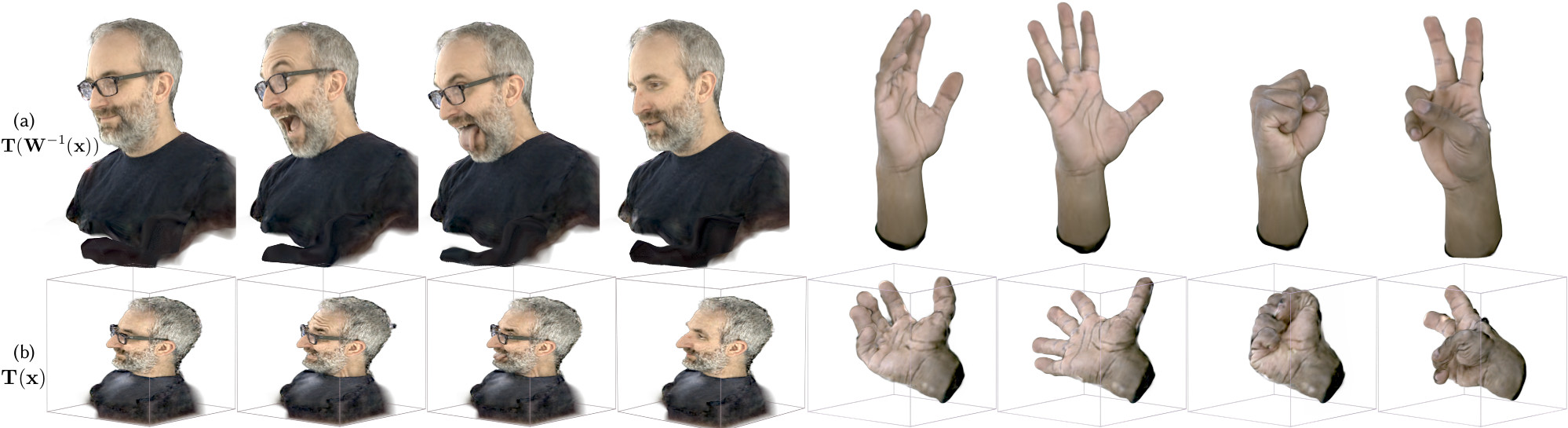}
    \caption{Warped and template volumes through time. This figure depicts several frames from two sequences, showing (a) the final warped volumes, and (b) the learned template volumes before warping. For the face sequence, most changes are explained via warping and the face remains static in template space. However, because tracking is not explicitly enforced, the network may learn to represent motion as different templates if it is more parsimonious to do so, as can be seen for finger motion in the hand sequence. Note also how in the template, space is allocated to regions requiring texture details while flat regions are compressed, like the phalanges and the forearm in the hand example.}
    \label{fig:templatemotion}
\end{figure*}

In this section, we show qualitative results on a number of different types of objects. We compare our renderings to held-out ground truth views and also demonstrate hybrid rendering with a mesh representation.

Fig.~\ref{fig:qualitative} shows renderings produced by our method compared to ground truth images for several validation (held-out) viewpoints. The renderings demonstrate that our method is able to model difficult phenomena like fuzz, smoke, and human skin and hair. The figure also shows typical artifacts produced by our reconstructions: typically, a very light smokey pattern is added which may be modeling view-dependent appearance for certain training cameras. Using more camera views tends to reduce all artifacts.

Fig.~\ref{fig:templatemotion} shows how the template volume changes through time across several frames, compared to the final warped volume that is rendered. Ideally, object motion should be represented entirely by the warp field. However, this does not always happen when representing such changes requires more resolution than is available (e.g., as would be necessary to cleanly separate the rim of the glasses from the side of the head) or because representing the warp field becomes too complex. 

\begin{figure*}
    \centering
    \includegraphics[width=1.0\textwidth]{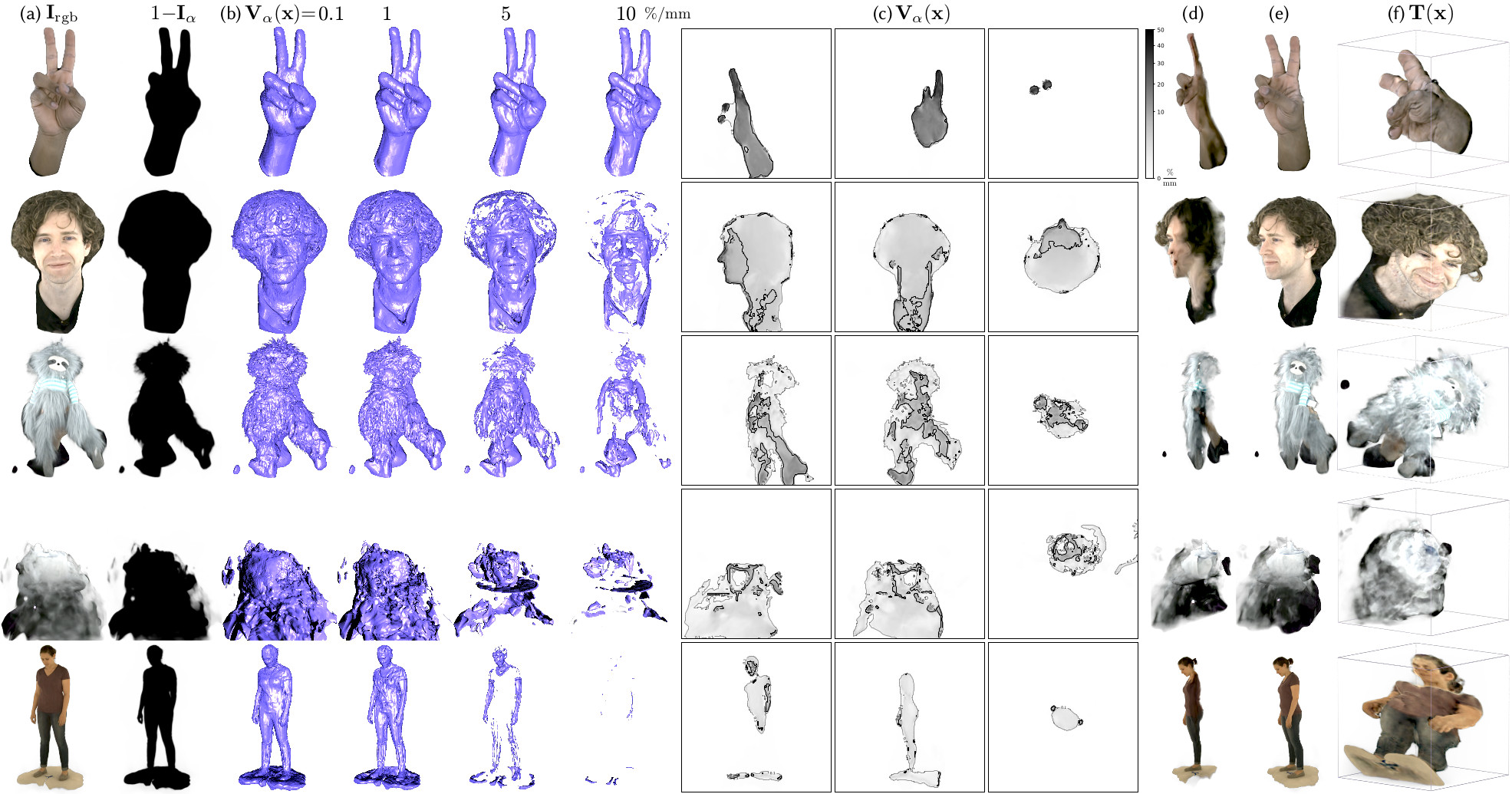}
    \caption{Learned reconstructions for several objects showing (a) rendered color and opacity, (b) isosurfaces of $\mathbf{V}_\alpha(\mathbf{x})$ for different levels of differential opacity, (c) contour plots of the differential opacity, (d) sliced render showing the interior of the volume at the x-y cut, (e) the complete render from the same viewpoint, and (f) the unwarped template.}
    \label{fig:isosurfaces}
\end{figure*}

Fig.~\ref{fig:isosurfaces} visualizes the learned RGB$\alpha$ volumes. While there is no explicit surface reconstruction objective in our method, we can visualize isosurfaces of constant opacity, shown in (b). Ideally, fully opaque surfaces such as the hand in the first row would be represented by delta functions in $\mathbf{V}_\alpha(\mathbf{x})$, but we find that the method trades off some opacity to better match reconstruction error in the training views. Note how translucent materials such as the glasses in the second row, or materials which appear translucent at coarse resolution, such as hair in the 3rd and 4th rows, are modeled with lower opacity values but retain a distinct structure particular to the object.



\subsection{Single Frame Estimation}

We evaluate the quality of our algorithm using only a single frame as input. In some ways this should make the problem easier as there is less information to represent within the model. On the other hand, our model is less able to exploit regularities and redundancies of motion. This experiment helps us disentangle those factors and determine the contribution of each component of the model.

To perform this experiment, we run our model on only a single frame. Although the encoder will produce a constant value, we keep the entire encoder-decoder network intact while training. We also compare to ``direct'' volume estimation, i.e., we directly estimate the template voxel volume $\mathbf{T}$ and warp values $\mathbf{W}^{-1}$ without the encoder-decoder network.

\begin{figure*}[t]
\centering
\includegraphics[width=1.0\textwidth]{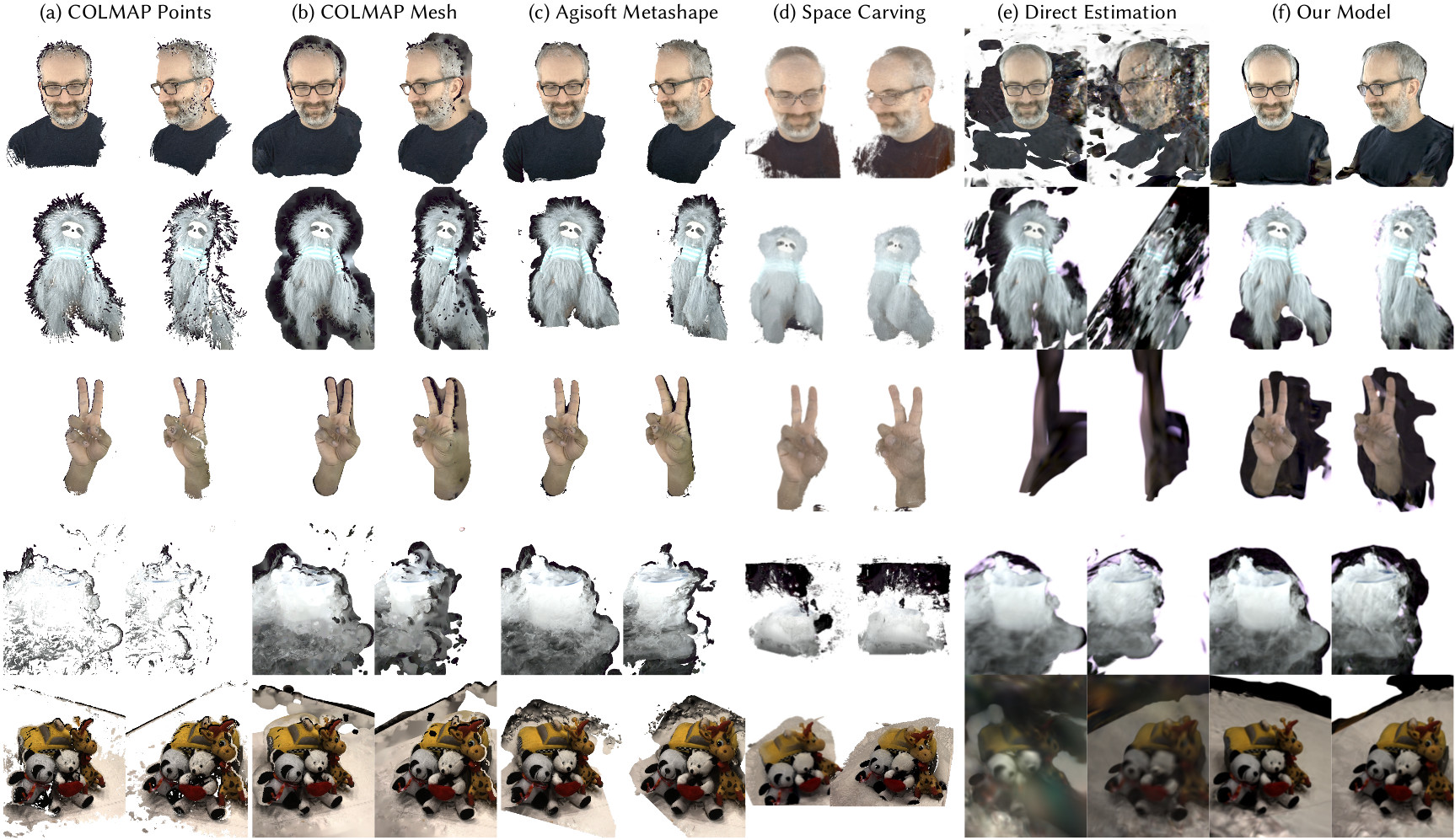}
\caption{Single frame estimation. In this experiment, we reconstruct only a single frame rather than a sequence. For comparison, from left to right, the first three columns show MVS-based reconstructions: (a) the point cloud recovered by COLMAP~\cite{Schoenberger:2016sfm,Schoenberger:2016mvs}, and (b) the mesh after Poisson reconstruction, and (c) the textured mesh recovered by Agisoft Metashape~\shortcite{Agisoft2019}. Column (d) shows a colored voxel occupancy reconstruction using space carving~\cite{Kutulakos:2000}, and column (e) shows ``direct'' estimation of the voxel grid (i.e., the $\mathbf{T}$ and $\mathbf{W}^{-1}$ are directly estimated instead of the result of an encoder-decoder network). Finally, (f) shows the results of our full pipeline trained only on a single frame. The results show that the encoder-decoder architecture facilitates recovering accurate estimates.}
\label{fig:1frame}
\end{figure*}

Fig.~\ref{fig:1frame} shows the results of the experiment on four objects we captured as well as one scene from a publicly available MVS dataset \cite{Aanaes:2016}. As shown, the ``direct'' voxel/warp estimation contains artifacts and incorrectly reconstructs the object. This experiment shows that the convolutional architecture provides a regularization that, even in the case of a single frame, allows us to accurately recover and re-render objects. Similar observations have been made in a recent work on deep image priors~\cite{Ulyanov:2018}.

For comparison, we show the same frame reconstructed as a mesh using a commercial multi-view stereo system~\cite{Agisoft2019} and the open-source multi-view stereo system COLMAP~\cite{Schoenberger:2016sfm,Schoenberger:2016mvs}, as well as a comparison to space carving~\cite{Kutulakos:2000}. Characteristically, the recovered resolution in the MVS texture map is greater than what we can achieve with volume reconstructions on current hardware. However, the mesh also shows artifacts in thin regions, like the frame of the glasses, and translucent regions, like the glass material and the hair at the top of the head, or the smoke in the 4th row. Space carving heavily relies on consistent appearance across views and struggles with translucent materials.

\subsection{Animating}

\begin{figure}[t]
    \begin{center}
    \includegraphics[width=1.0\columnwidth]{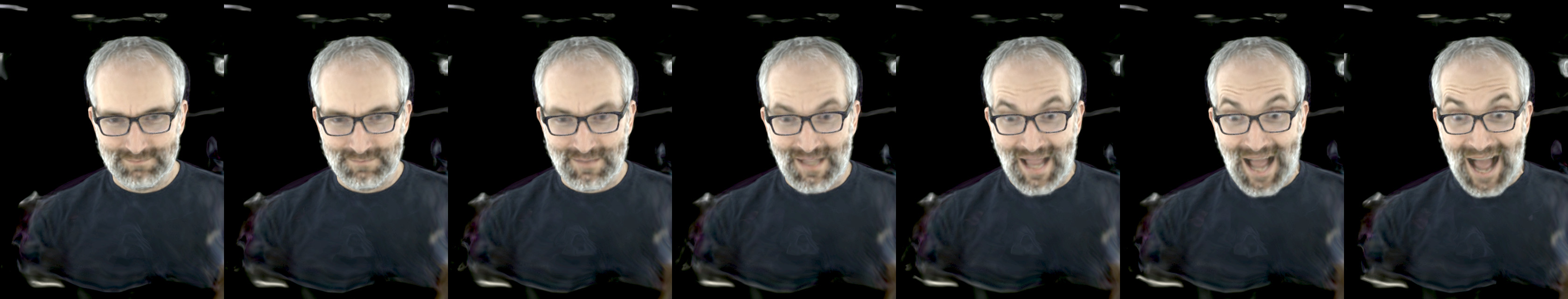}

    \includegraphics[width=1.0\columnwidth]{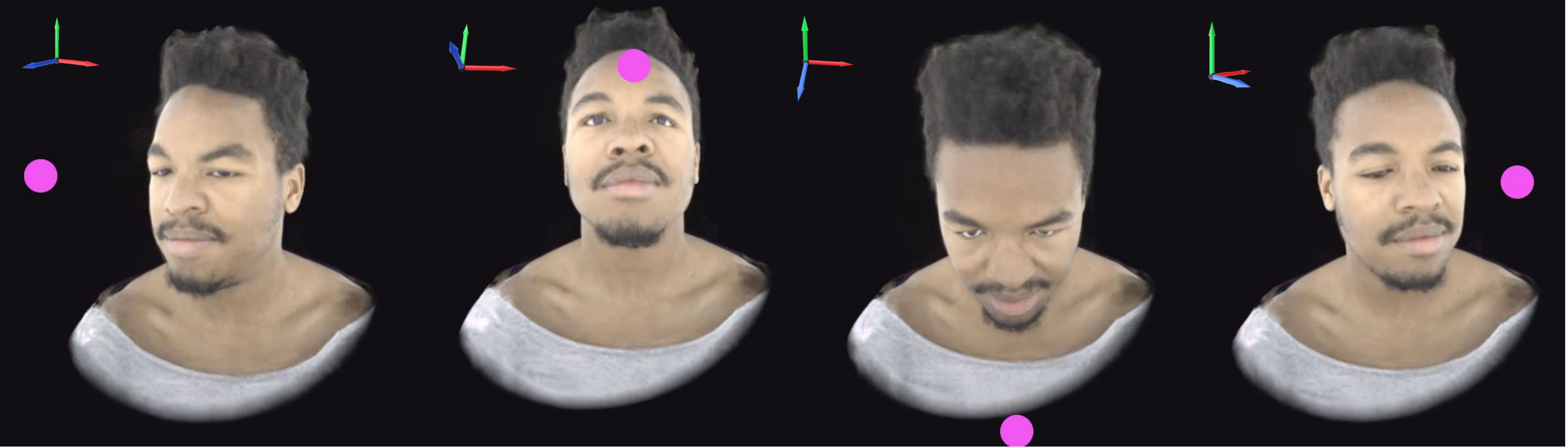}
    \end{center}
\caption{Novel content generation. First row: Interpolation in latent space. Leftmost and rightmost frames are reconstructed from real image encodings, intermediate frames are reconstructed from linear interpolation between the left and right codes. Second row: real-time avatar driving based on user input. The magenta dot represents the position of the user's hand, which the avatar's head turns to track.}
\label{fig:driving}
\end{figure}

With the proposed reconstruction method, we can playback and re-render the captured data from many different angles. We not only want to extrapolate in viewpoint, but also in the content of the performance. Our latent variable model allows us to create new sequences of content by modifying the latent variables.

Fig.~\ref{fig:driving} shows two examples of content modification by interpolating latent codes and by changing the conditioning variable based on user input. Our latent space interpolation shows that the encoder network learns a compact representation of the scene. In the second example, we condition the decoder on the head pose of the subject, allowing us to create novel sequences in real time.






\begin{figure}[t]
    \begin{center}
	    \includegraphics[width=0.24\columnwidth]{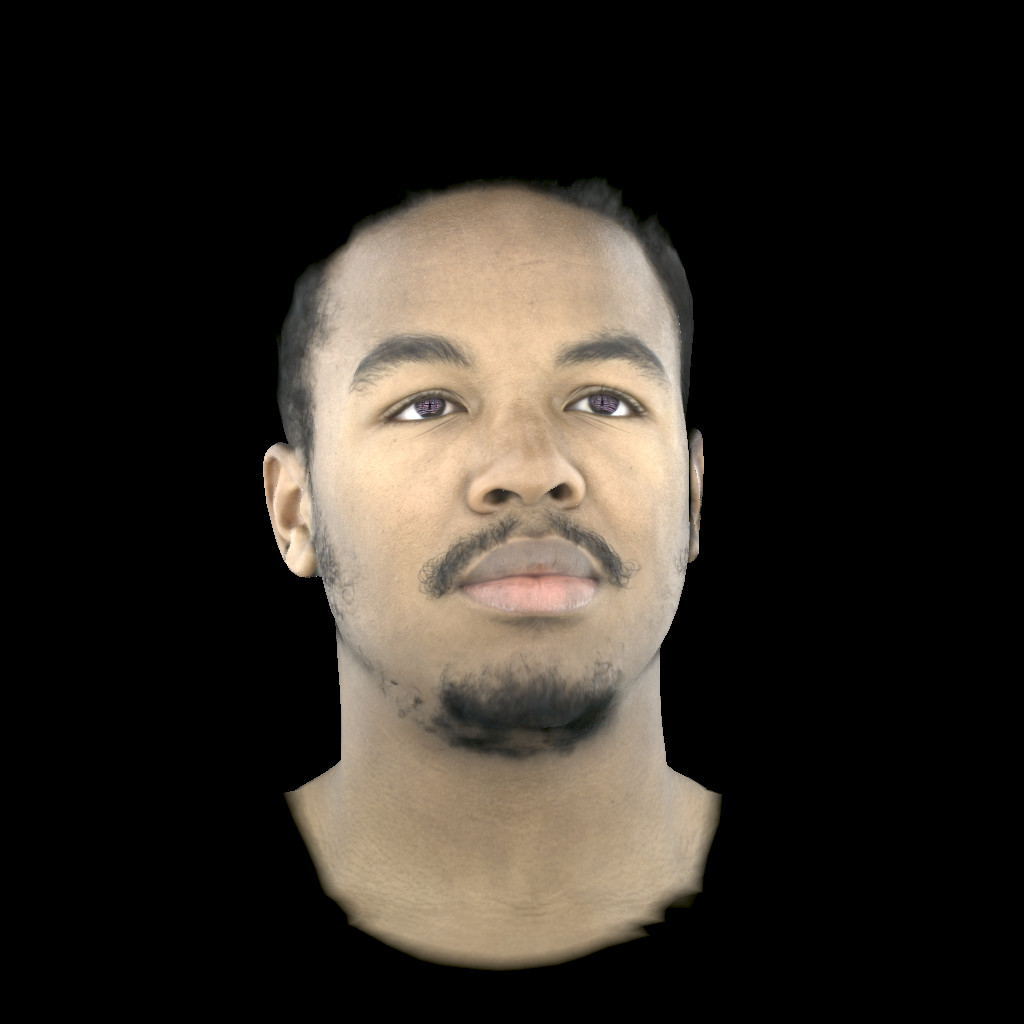}
	    \includegraphics[width=0.24\columnwidth]{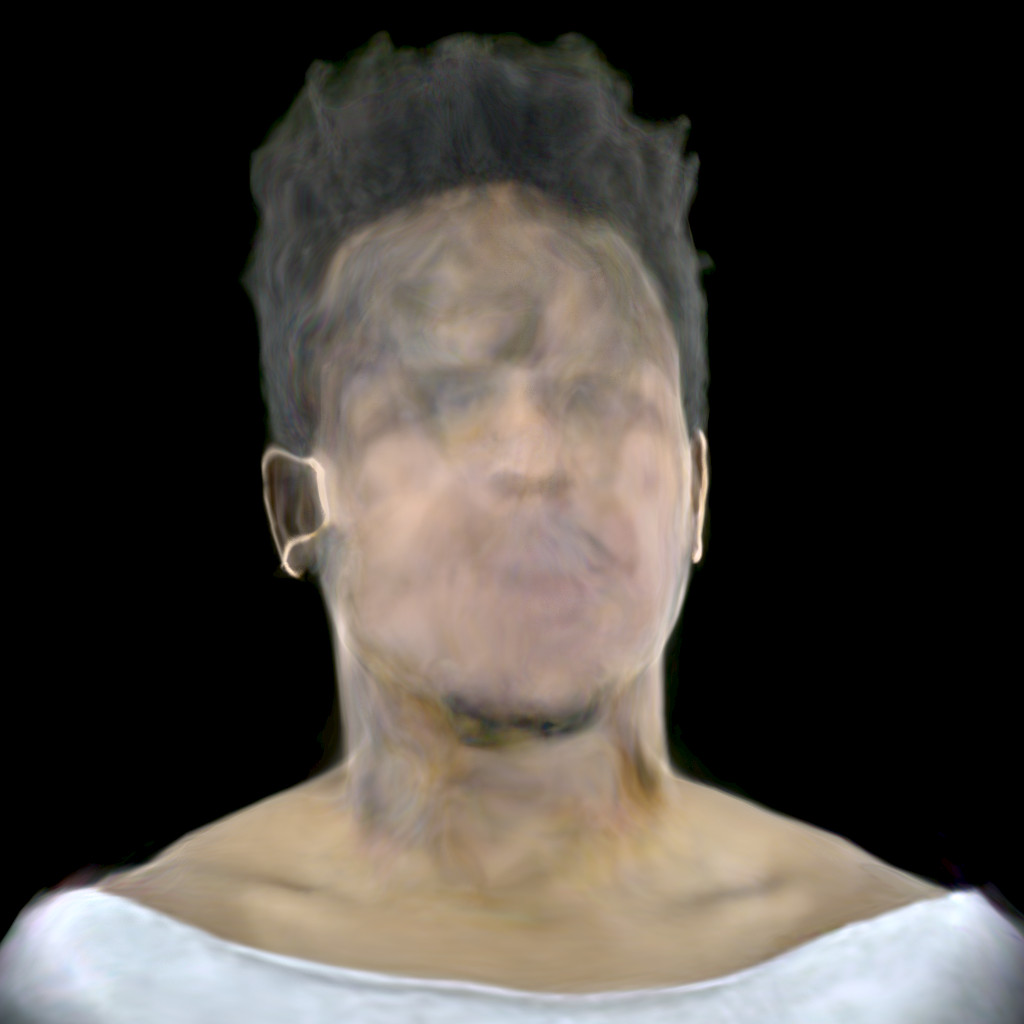}
	    \includegraphics[width=0.24\columnwidth]{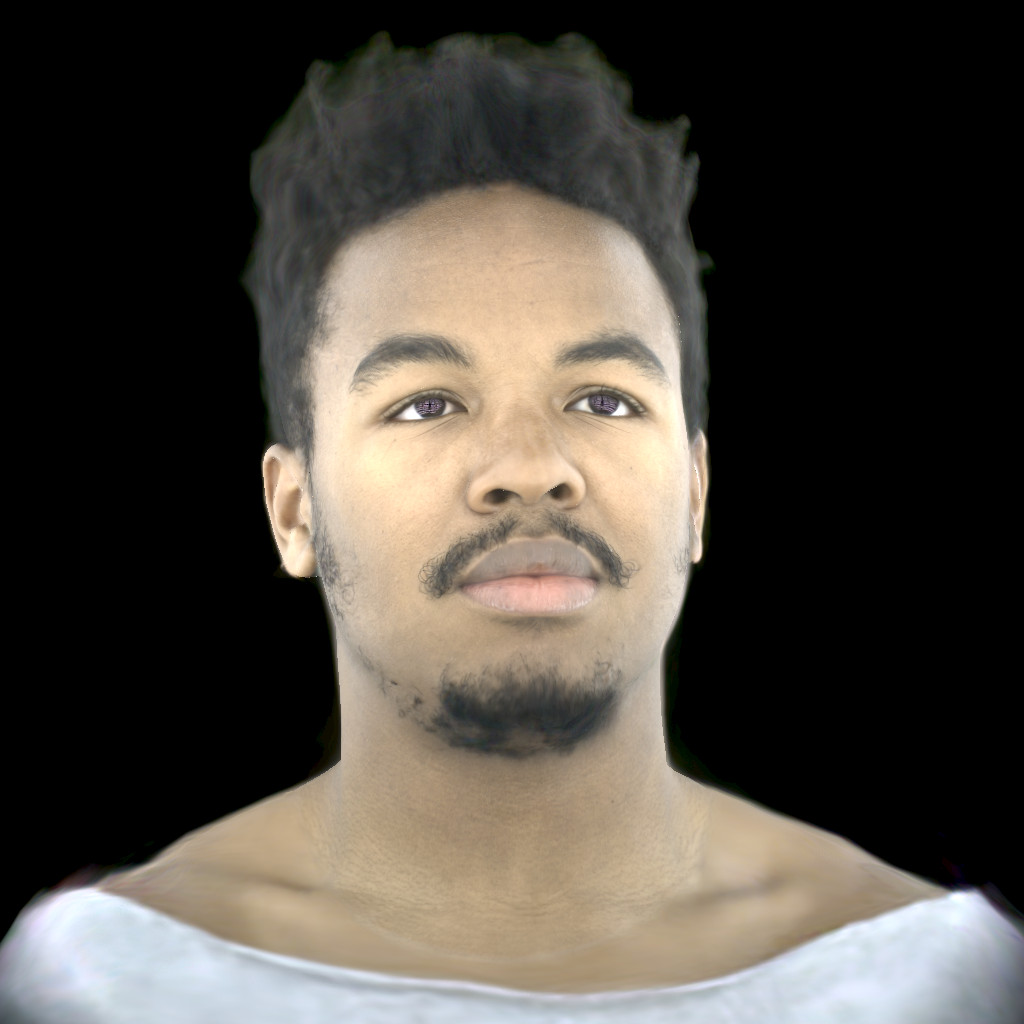}
	    \includegraphics[width=0.24\columnwidth]{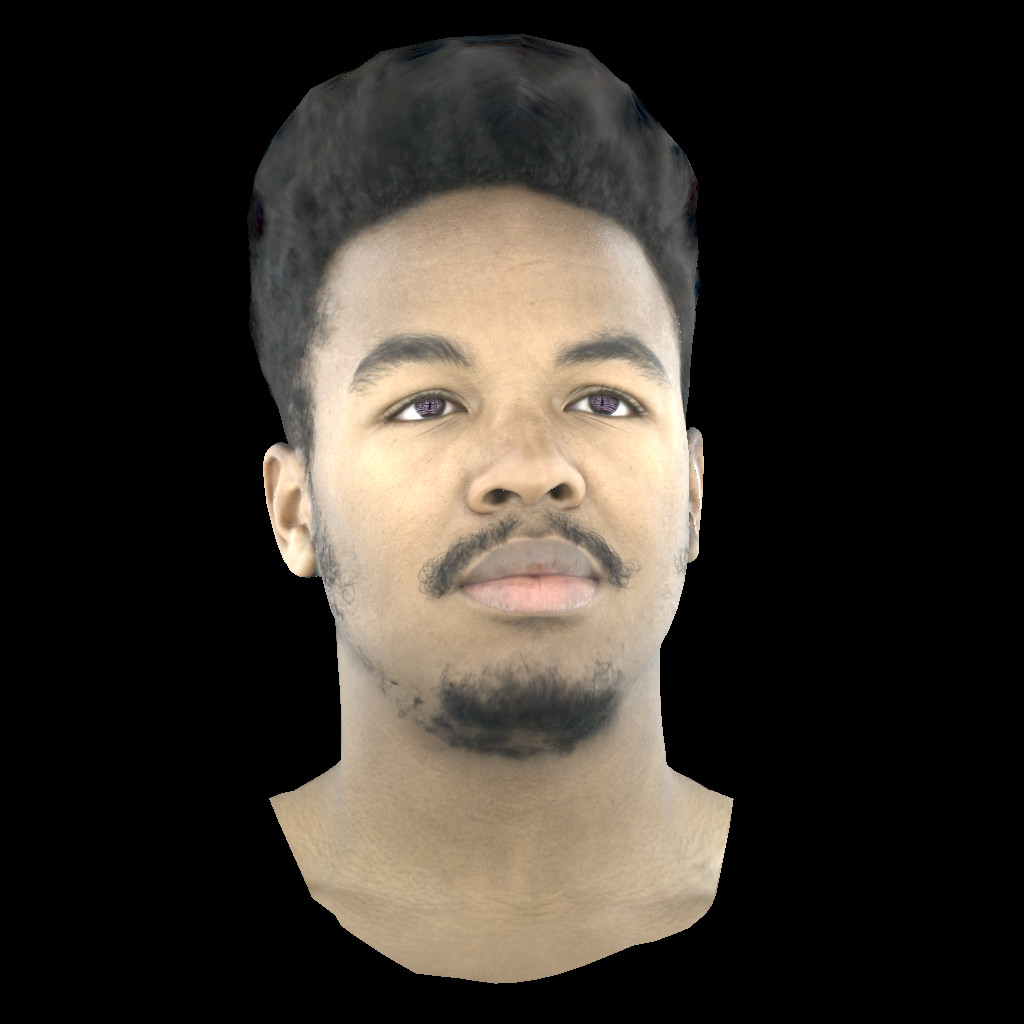}
	\end{center}
	\begin{center}
	    \includegraphics[width=0.24\columnwidth]{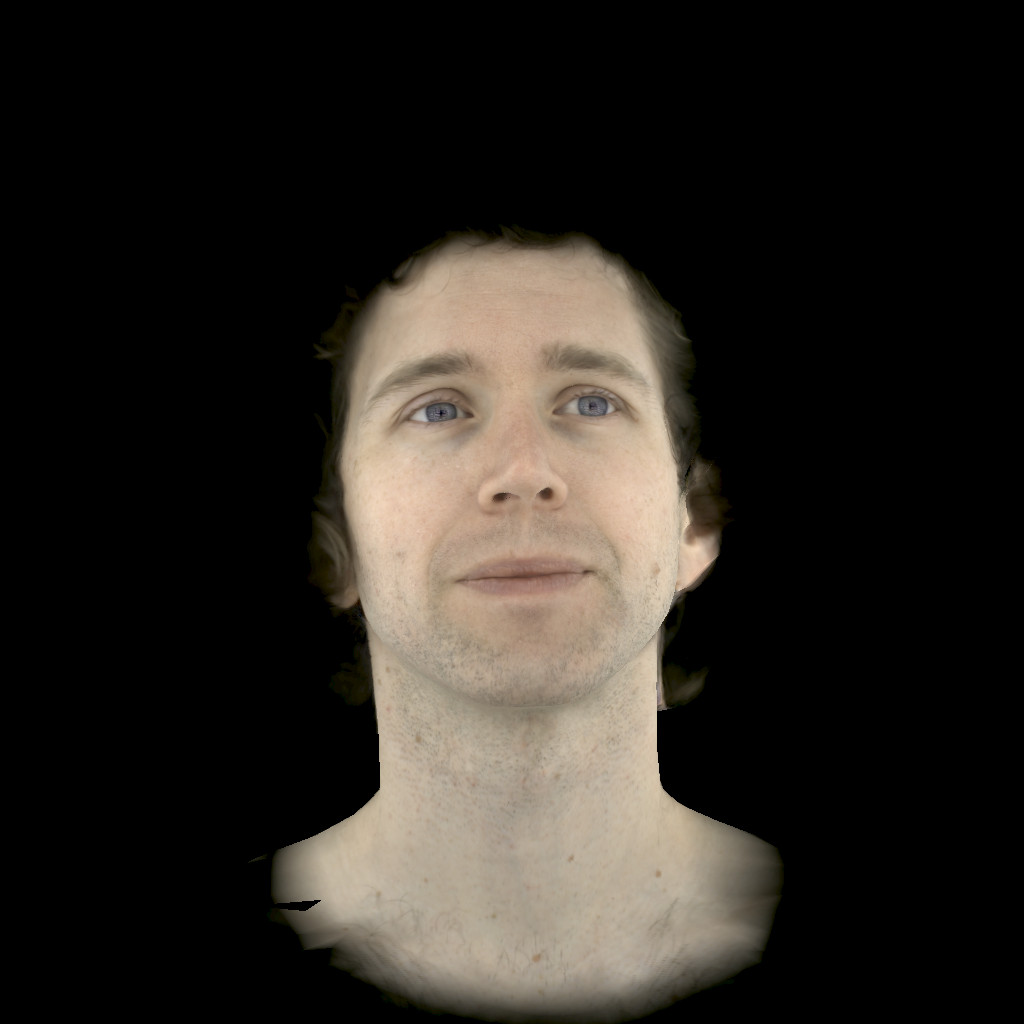}
	    \includegraphics[width=0.24\columnwidth]{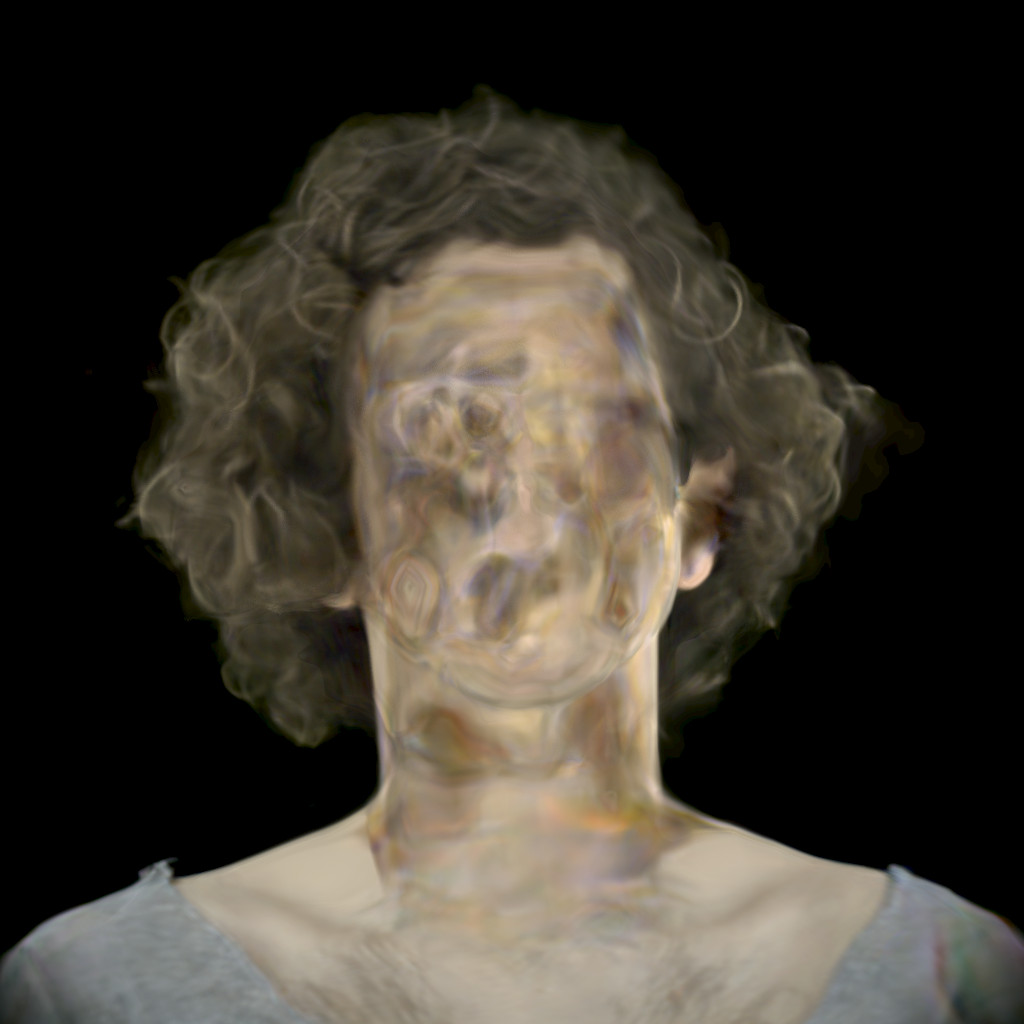}
	    \includegraphics[width=0.24\columnwidth]{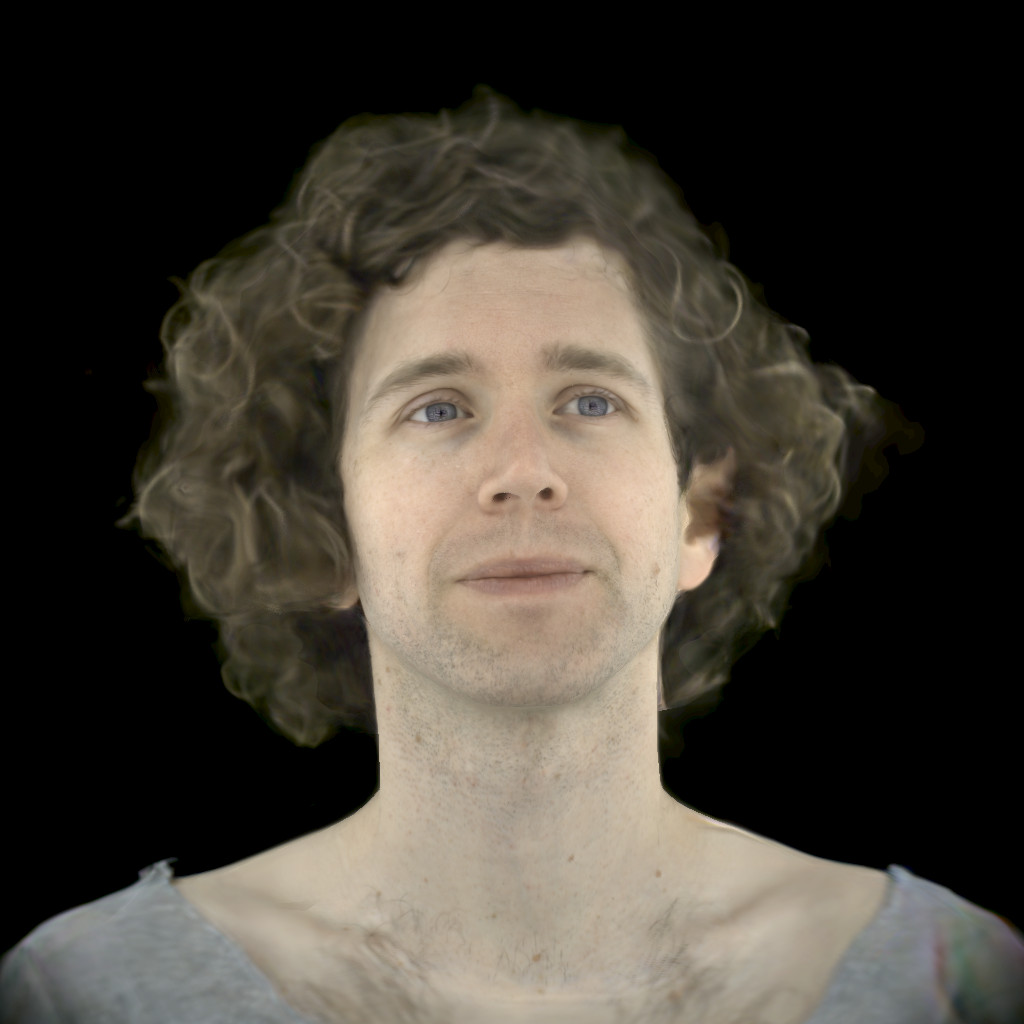}
	    \includegraphics[width=0.24\columnwidth]{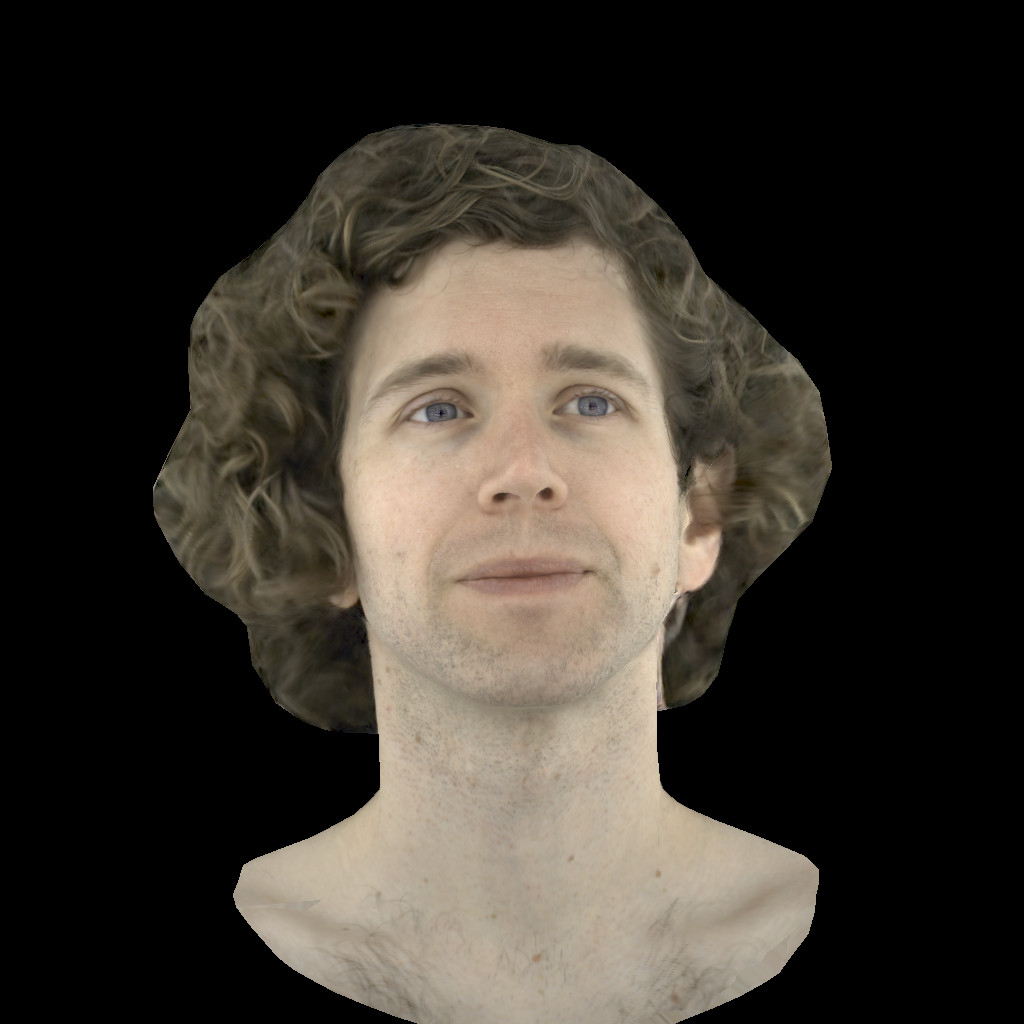}
    \end{center}
\caption{Combining mesh and volumetric representations (images best viewed at high-resolution). Given an initial textured mesh model, all rays which intersect the mesh have $t_{\mathrm{max}}$ set to the depth of the mesh along each ray. Rays which terminate at $t_{\mathrm{max}}$ accumulate color from the mesh based on remaining opacity. From left to right: masked mesh (placed into voxel volume during learning), voxel representation, hybrid rendering, mesh-only reconstruction. The learned voxel representation avoids occluding the mesh to achieve a higher-quality reconstruction.}
\label{fig:meshvoxel}
\end{figure}

\label{subsec:meshvoxel}
Fig.~\ref{fig:meshvoxel} shows the results of combining a textured mesh representation with our voxel representation. Textured meshes can efficiently and accurately represent fine detail in regions of the face like the skin and eyes while the voxel representation excels at modeling hair. For a mesh model we use the Deep Appearance Model of \mbox{\citet{Lombardi:2018} trained} to reconstruct the same face we used to train our volume encoder/decoder network.

\begin{figure}
    \centering
    \includegraphics[width=1.0\columnwidth]{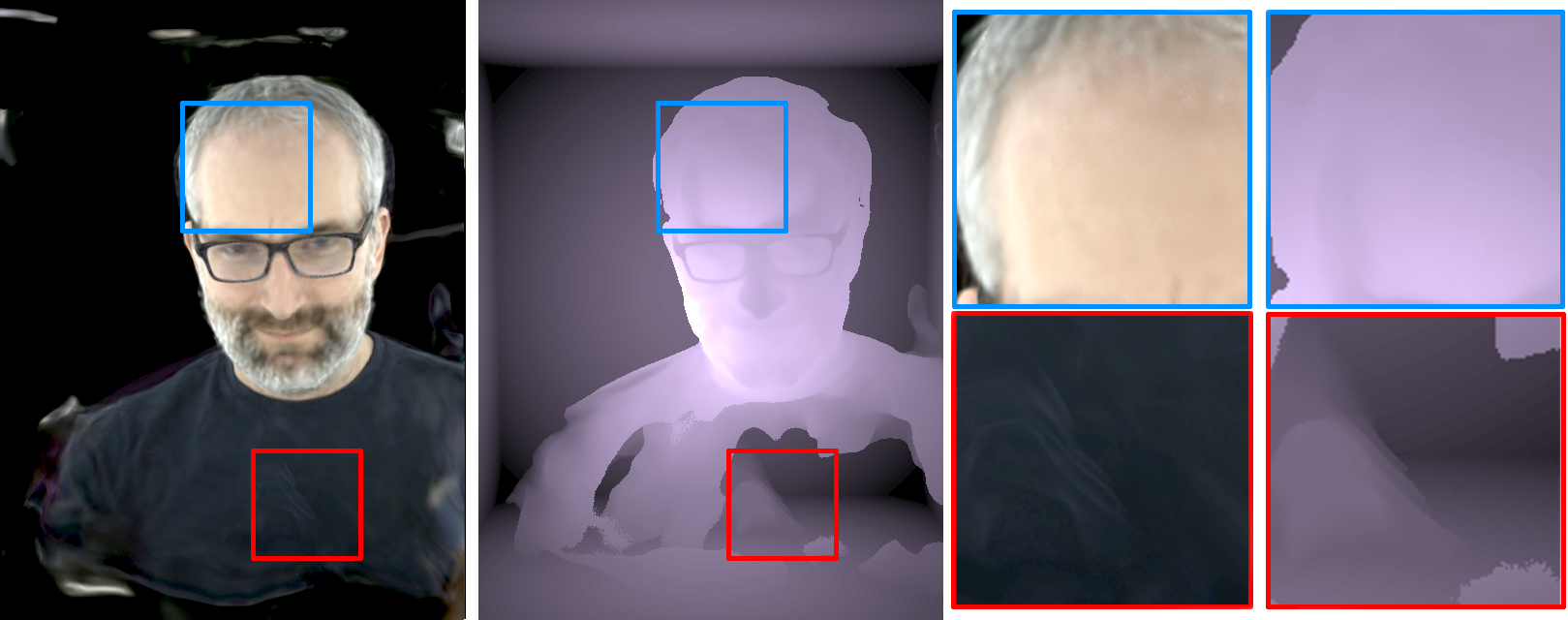}
    \caption{Depth map showing where raymarching terminates. The hole in the chest indicates that those rays passed through the entire volume and only terminated at the background. Since the chest area has limited texture variation and is a similar color to the background, this artifact does not greatly affect reconstruction error even when viewed from novel viewpoints.}
    \label{fig:depth_map}
\end{figure}

\section{Discussion} \label{sec:discussion}

In this paper, we presented a method for modeling objects and scenes with a semi-transparent volume representation that we learn end-to-end from multi-view RGB images. We showed that our method can convincingly reconstruct challenging objects such as moving hair, fuzzy toys, and smoke. Our method requires no explicit tracking, and can be run in real time alongside traditional triangle rasterization.

One limitation of our method is that given a surface with limited texture, our estimated volume may represent that surface as transparent and place its color in the background, so long as doing so does not cause otherwise occluded surfaces to appear. This is a challenge that affects traditional 3D reconstruction methods as well. With our method, however, the reconstruction degrades gracefully and still produces perceptually-pleasing results thanks to our image-space loss function. Fig.~\ref{fig:depth_map} shows an example of this via a depth map computed as the distance each ray travels before saturating or hitting the bounding box. In a more practical setting, this could be addressed simply by capturing the sequence with a bright background such as a green screen.

Although our method can handle transparent objects, like plastic bottles, the method doesn't currently consider refractive surfaces. We believe the approach can be extended to model refraction and even reflection, and we leave that to future work. Our model can represent dull specular highlights through view conditioning but high-frequency specular highlights are not correctly represented.

The latent space is the feature enabling us to generate dynamic content, but we do not explicitly model any temporal dynamics. This is not a problem for playback, since the playback sequence implicitly encodes the same temporal dynamics as the recording. It is also not a problem when driving the representation from user input, so long as that user input has reasonable temporal dynamics of its own. However, if we traverse the latent space in some manner not guided by temporal information, we may generate sequences which, while visually accurate, do not represent real behaviors of the object we modeled.

Volumetric representations typically suffer from limited resolution due to the cubic relationship between resolution and memory requirement. In this work, we showed some ways to increase the effective resolution without simply increasing the voxel grid resolution by using warping fields. We believe that we can further improve this approach to achieve a level of fidelity and resolution previously only achievable with traditional textured mesh surfaces.

%
%
%
%


\bibliographystyle{ACM-Reference-Format}
\bibliography{main}

\end{document}